\documentclass[11pt,a4paper]{article}%
\usepackage{amsfonts}
\usepackage{amsmath}
\usepackage{amssymb}
\usepackage{graphicx}%
\providecommand{\U}[1]{\protect\rule{.1in}{.1in}}

\begin{document}

\title{\begin{flushright}
\vspace{-2cm}
\small{Bicocca-FT-08-02}\\
\small{NSF-KITP-07-212}\\
\end{flushright}
\vspace{1cm}
Eikonal Methods in AdS/CFT:\\BFKL Pomeron at Weak Coupling}
\author{Lorenzo Cornalba$^{a}$, Miguel S. Costa$^{b}$, Jo\~ao Penedones$^{b,c}$ \medskip\\$^{a}$ Centro Studi e Ricerche E. Fermi, Compendio Viminale, I-00184, Roma\\Universit\`{a} di Milano-Bicocca and INFN, sezione di Milano-Bicocca\\Piazza della Scienza 3, I--20126 Milano, Italy \medskip\\$^{b}$ Departamento de F\'{\i}sica e Centro de F\'{\i}sica do Porto,\\Faculdade de Ci\^{e}ncias da Universidade do Porto,\\Rua do Campo Alegre 687, 4169--007 Porto, Portugal \medskip\\$^{c}$ Kavli Institute for Theoretical Physics\\University of California, Santa Barbara, CA 93106-4030, USA\medskip\\{\small {\texttt{Lorenzo.Cornalba@mib.infn.it, miguelc@fc.up.pt,
penedon@kitp.ucsb.edu}}}}
\date{}
\maketitle

\begin{abstract}
We consider correlators of $\mathcal{N}=4$ super Yang Mills of the form
$\mathcal{A\sim}\left\langle \mathcal{O}_{1}\mathcal{O}_{2}\mathcal{O}%
_{1}^{\star}\mathcal{O}_{2}^{\star}\right\rangle $, where the operators
$\mathcal{O}_{1}$ and $\mathcal{O}_{2}$ are scalar primaries. In particular, we
analyze this correlator in the planar limit and
in a Lorentzian regime corresponding to high energy interactions in AdS. The
planar amplitude is dominated by a Regge pole whose
nature varies as a function of the 't Hooft coupling $g^{2}$. At large $g$, the
pole corresponds to graviton exchange in AdS, whereas at weak $g$, the pole is
that of the hard perturbative BFKL pomeron. We concentrate on the weak
coupling regime and analyze pomeron exchange directly in position space. The
analysis relies heavily on the conformal symmetry of the transverse space
$\mathbb{E}^{2}$ and of its holographic dual hyperbolic space $\mathrm{H}_{3}%
$, describing  with an unified language, both the weak and strong
't Hooft coupling regimes. In particular, the form of the impact factors
is highly constrained in position space by conformal invariance. Finally, the
analysis suggests a possible AdS eikonal resummation of multi-pomeron
exchanges implementing AdS unitarity, which differs from the usual
$4$--dimensional eikonal exponentiation. Relations to violations of
$4$--dimensional unitarity at high energy and to the onset of nonlinear
effects and gluon saturation become immediate questions for future research.

\end{abstract}

\section{Introduction}

\begin{figure}
\begin{center}
\includegraphics[
height=1.3474in
]{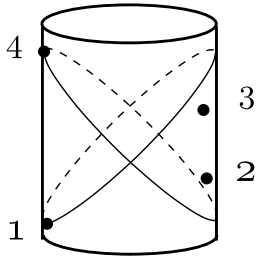}
\end{center}
\caption{CFT points $\mathbf{x}_i$ on the boundary of global AdS. Shown is the 
relevant Lorentzian
kinematics, with $\mathbf{x}_{4}$ in the future of $\mathbf{x}_{1}$,
$\mathbf{x}_{3}$ in the future of $\mathbf{x}_{2}$, and with the pairs
$\mathbf{x}_{1}$, $\mathbf{x}_{2}$ and $\mathbf{x}_{3}$, $\mathbf{x}_{4}$
spacelike related. This choice corresponds to a $2$ to $2$ interaction in the bulk of AdS.}
\label{fig8}
\end{figure}

String interactions in flat space are dominated, at tree level and in the
eikonal regime $s\gg\left\vert t\right\vert $, by a leading Regge pole
associated to the exchange of string excitations of increasing spin,
starting with the massless graviton. The same behavior is expected for high energy
interactions of strings in AdS \cite{PS1,Paper1, Paper2, Paper3, PS2, PS3, Paper4}. 
In this case, the flat space $S$--matrix is substituted by correlators
in the dual conformal field theory, and the analogous of a $2$ to $2$
scattering amplitude is given by CFT correlators of the form%
\begin{equation}
\mathcal{A}\sim\left\langle \mathcal{O}_{1}\left(  \mathbf{x}_{1}\right)
\mathcal{O}_{2}\left(  \mathbf{x}_{2}\right)  \mathcal{O}_{1}^{\star}\left(
\mathbf{x}_{3}\right)  \mathcal{O}_{2}^{\star}\left(  \mathbf{x}_{4}\right)
\right\rangle ~. \label{in1}%
\end{equation}
CFT positions $\mathbf{x}_{i}$ play the role of momenta in AdS, with the
analogous of a scattering process achieved by choosing  Lorentzian
kinematics with $\mathbf{x}_{4}$ in the future of $\mathbf{x}_{1}$,
$\mathbf{x}_{3}$ in the future of $\mathbf{x}_{2}$, and with the pairs
$\mathbf{x}_{1}$, $\mathbf{x}_{2}$ and $\mathbf{x}_{3}$, $\mathbf{x}_{4}$
spacelike related \cite{Paper1, Paper2, Paper3}, as shown in figure \ref{fig8}.
The relevant AdS eikonal regime is then obtained by sending $\left(
\mathbf{x}_{3}-\mathbf{x}_{1}\right)  ^{2},$ $\left(  \mathbf{x}%
_{4}-\mathbf{x}_{2}\right)  ^{2}\rightarrow0$. Contrary to a Euclidean
configuration, the amplitude $\mathcal{A}$ is not dominated in this limit by
the OPE, but by the exchange of operators of maximal spin \cite{Paper2}, as in flat space.
Moreover, whenever the spin of the exchanged operators is unbounded, one must
use Regge techniques, as discussed in detail in \cite{Paper4, PS3}.

We shall focus our attention on the canonical example of Type
IIB strings on AdS$_{5}\times$S$_{5}$, dual to $\mathrm{SU}\left(  N\right)
,$ $\mathcal{N}=4$ super Yang--Mills (SYM) \cite{AdsCFT}. The planar contribution $N^{-2}%
\mathcal{A}_{\mathrm{planar}}$ to the full amplitude $\mathcal{A}$ corresponds
to tree level string interactions in AdS and will be dominated by a Regge pole
whose trajectory $j\left(  \nu,g\right)  $ will depend on the 't Hooft
coupling $g^{2}=g_{\mathrm{YM}}^{2}N$ of the Yang--Mills theory, or dually on
the AdS radius $\ell$ in units of string length $\sqrt{\alpha^{\prime}}$.
Moreover, the trajectory $j\left(  \nu,g\right)  $ depends, as in flat space,
on the transverse momentum transfer $\sqrt{-t}=\nu/\ell$. At large coupling
$g=\ell^{2}/\alpha^{\prime}$, strings move almost in flat space, and the Regge
spin is given essentially by the flat space trajectory $2+\alpha^{\prime}t/2$
so that \cite{PS1, Paper4}%
\[
j\left(  \nu,g\right)  =2-\frac{\nu^{2}}{2g}-\frac{2}{g}-\cdots
~,~\ \ \ \ \ \ \ \ \ \ \ \ \ \ \ \ \left(  g\rightarrow\infty\right)\ .
\]
Only the third term is not determined by the flat space limit, since it
vanishes for $\ell\rightarrow\infty$. It is fixed, however, by the
requirement that the graviton is massless in AdS for any value of $g$, which
translates to $j\left(  \pm2i,g\right)  =2$, as shown in \cite{Paper4}. This
implies that, as we decrease the radius of AdS,
the intercept $j\left(  0,g\right)  =2-2/g-\cdots$
decreases from the flat space result.

This paper is concerned, on the other hand, with the leading Regge pole of
$\mathcal{N}=4$ SYM at weak 't Hooft coupling. The high energy behavior of
SYM, when analyzed in momentum space in four dimensions and in the high energy
regime $s\gg\left\vert t\right\vert $, is dominated by the exchange of a
single perturbative BFKL pomeron \cite{BFKL, Lipatov, LipatovRev}. To leading
logarithmic order, the BFKL\ pomeron is independent of the underlying
supersymmetry, and dominates high energy interactions as in conventional QCD.
At leading order in $g^{2}$, the pomeron is nothing but a pair of gluons in a
color singlet state of effective spin $1$. Moreover, the leading corrections
in $g$ modify this trajectory to%
\[
j\left(  \nu,g\right)  =1+\frac{g^{2}}{4\pi^{2}}\left(  2\Psi\left(  1\right)
-\Psi\left(  \frac{1+i\nu}{2}\right)  -\Psi\left(  \frac{1-i\nu}{2}\right)
\right)  +\cdots~.
\]
Note that the leading intercept $j\left(  0,g\right)  =1+g^{2}\ln2/\pi
^{2}+\cdots$ increases for small $g^{2}$, justifying the conjecture
\cite{PS1}\ that the pomeron trajectory is nothing but the leading string
trajectory at weak coupling, corresponding to string exchange in a highly
curved AdS spacetime.

The usual treatment of BFKL  pomeron exchange is conducted in $4$--dimensional
momentum space. More precisely, external scattering states are chosen to be
momentum eigenstates with appropriate kinematics, whereas the internal pomeron
propagator is best described in position space on the space $\mathbb{E}^{2}$
transverse to the interaction \cite{Lipatov}. However, as discussed
above, the more appropriate way to analyze SYM correlators, in view of the
AdS/CFT duality, is to consider them as \textit{\textquotedblleft}%
$S$\textit{--matrix elements\textquotedblright} of interactions in AdS, with
CFT positions playing the role of AdS momenta. It is then natural to
reconsider the BFKL analysis with external states labeled by positions, in the
kinematical limit described at the beginning of this introduction. We shall
address this issue, sharpening the conjectured duality between
pomeron exchange and string exchange in AdS. More precisely, we analyze the
couplings of external states to the BFKL pomeron -- the so--called impact
factors -- in position space, heavily using the conformal invariance
$SO\left(  3,1\right)  $ of the transverse conformal space $\mathbb{E}^{2}$
and of its holographic dual hyperbolic $3$--space $\mathrm{H}_{3}$. The
formalism allows us to describe, in a unified and coherent fashion, the Regge
pole exchange at weak coupling as well as at strong coupling.

We shall work mostly with a specific simple example, where the operators
$\mathcal{O}_{1}$ and $\mathcal{O}_{2}$ are given by the chiral primaries
$\mathrm{Tr}\left(  Z^{2}\right)  $ and $\mathrm{Tr}\left(  W^{2}\right)  $,
with $Z$ and $W$ two of the complex adjoint scalar fields of $\mathcal{N}=4$ SYM.
The correlator (\ref{in1}) is known both at weak coupling \cite{FourPT}\ at
order $g^{4}$, as well as at strong coupling using the AdS/CFT duality
\cite{Strong}, and it is therefore a good example to describe the general
theory. In section \ref{SecG}, after reviewing some facts on Regge theory in
CFT's \cite{Paper4}, we summarize the general results of the paper. Sections
\ref{SecBFKL} and \ref{SecN4} are then devoted to the proof of these results.
More precisely, in section \ref{SecBFKL} we discuss the general BFKL formalism
in position space for generic scalar operators $\mathcal{O}_{1}$ and
$\mathcal{O}_{2}$, whereas in section \ref{SecN4} we specialize to the
operators $\mathrm{Tr}\left(  Z^{2}\right)  $ and $~\mathrm{Tr}\left(W^{2}\right)  $, 
deriving their impact factors in position space. Since section \ref{SecG}
contains a summary of our main findings, as well as open questions, we
found it redundant to include a concluding section.

The present paper is focused mostly on the analysis of the planar limit
$N^{-2}\mathcal{A}_{\mathrm{planar}}$ of the correlator (\ref{in1}), which
corresponds to the exchange of a single pomeron. On the other hand, 
due to the raising intercept $j(0,g)>1$, a single pomeron
exchange generates  a total cross section that grows with energy and
inevitably violates unitarity bounds in four dimensions \cite{BK,Iancu, HIM}. 
At weak coupling, it is well known that this problem cannot be
cured uniquely by eikonalizing the pomeron exchange, but one must also consider
non--linear pomeron interactions which tame the high energy growth and restore
unitarity. In the context of hadronic interactions, this corresponds to the
saturation of the hadron gluon transverse density at small values of Bjorken
$x$ and is quite relevant to experimentally accessible regimes in deep
inelastic scattering experiments \cite {BK, Iancu, Mueller}. Our position space
formalism, on the other hand, is related from the start to interactions in
AdS and, in fact, admits an eikonalization with respect to geodesic motion in
five dimensions \cite{Paper3, Paper4}. Moreover, the AdS eikonal is clearly valid at
strong 't Hooft coupling, where, for a large range of AdS impact parameters,
the phase shift is of order one and needs to be eikonalized even though one is
quite far from the critical impact parameter where non--linear gravitational
effects start to become important and drive black hole formation. It is then
tempting to speculate that, even at weak coupling, the $5$--dimensional AdS
eikonal resummation is valid in some range of the kinematical parameters, and
is relevant for the physics of high energy scattering before the onset of
gluon saturation. These issues, as well as the fascinating relation between
gluon saturation and black hole formation where non linear effects become
important \cite{Nastase, AG1, AG2}, could lead to a possible experimental
observation of the gauge/gravity correspondence and will be the subject of our
future investigations \cite{Paper6}.

\section{General Results\label{SecG}}

\subsection{Review of Regge theory for CFT's\label{SecGeneral}}

\begin{figure}[ptb]
\begin{center}
\includegraphics[
height=1.8474in
]{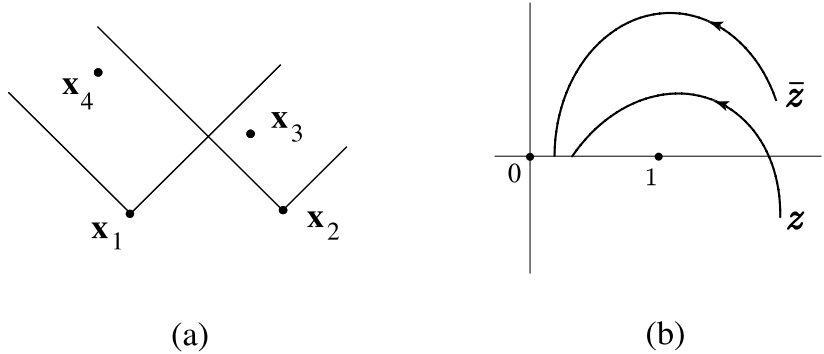}
\end{center}
\caption{(a) The kinematics (\ref{in100}) and (\ref{in100BIS})
for the Lorentzian amplitude $\hat{\mathcal{A}}$.
(b) For this kinematics we show the relevant analytic
continuation in $z,\bar{z}$ for $\hat{\mathcal{A}}$,
starting from the Euclidean amplitude
$\mathcal{A}$ with $\bar{z}=z^{\star}$. }
\label{fig7}
\end{figure}

We consider a $4$--dimensional conformal field theory defined on
Minkowski space $\mathbb{M}^{4}$. We parameterize points $\mathbf{x}%
\in\mathbb{M}^{4}$ with two light--cone coordinates $x^{+},x^{-}$ and with a
point $x$ in the Euclidean transverse space $\mathbb{E}^{2}$, and we choose
the metric $-dx^{+}dx^{-}+dx\cdot dx$. We will be interested in the analysis
of the correlator%
\begin{equation}
\left\langle \mathcal{O}_{1}\left(  \mathbf{x}_{1}\right)  \mathcal{O}%
_{2}\left(  \mathbf{x}_{2}\right)  \mathcal{O}_{1}^{\star}\left(
\mathbf{x}_{3}\right)  \mathcal{O}_{2}^{\star}\left(  \mathbf{x}_{4}\right)
\right\rangle ~, \label{in000}%
\end{equation}
where $\mathcal{O}_{1}$ and $\mathcal{O}_{2}$ are scalar primary operators of
dimension $\Delta_{1}$ and $\Delta_{2}$, respectively. By conformal
invariance, the above correlator can be expressed as%
\[
\frac{1}{\left(  \mathbf{x}_{1}-\mathbf{x}_{3}\right)  ^{2\Delta_{1}}\left(
\mathbf{x}_{2}-\mathbf{x}_{4}\right)  ^{2\Delta_{2}}}~\mathcal{A}\left(
z,\bar{z}\right)  ~,
\]
where the reduced amplitude $\mathcal{A}$ depends on the cross--ratios
$z,\bar{z}$ defined by \cite{Osborn}%
\begin{align*}
z\bar{z}  &  =\frac{\left(  \mathbf{x}_{1}-\mathbf{x}_{3}\right)  ^{2}\left(
\mathbf{x}_{2}-\mathbf{x}_{4}\right)  ^{2}}{\left(  \mathbf{x}_{1}%
-\mathbf{x}_{2}\right)  ^{2}\left(  \mathbf{x}_{3}-\mathbf{x}_{4}\right)
^{2}}~,\newline\\
\left(  1-z\right)  \left(  1-\bar{z}\right)   &  =\frac{\left(
\mathbf{x}_{1}-\mathbf{x}_{4}\right)  ^{2}\left(  \mathbf{x}_{2}%
-\mathbf{x}_{3}\right)  ^{2}}{\left(  \mathbf{x}_{1}-\mathbf{x}_{2}\right)
^{2}\left(  \mathbf{x}_{3}-\mathbf{x}_{4}\right)  ^{2}}~~.
\end{align*}
The reduced amplitude is originally defined for Euclidean configurations with
$\left(  \mathbf{x}_{i}-\mathbf{x}_{j}\right)  ^{2}>0$ and $\bar{z}=z^{\star}%
$, where it coincides with the amplitude of the Euclidean continuation of the
CFT at hand. On the other hand, here we are interested in
intrinsically Lorentzian configurations, with%
\begin{align}
&  \mathbf{x}_{4}\text{ in the future of }\mathbf{x}_{1}~,\nonumber\\
&  \mathbf{x}_{3}\text{ in the future of }\mathbf{x}_{2}~,\label{in100}\\
&  \left(  \mathbf{x}_{i}-\mathbf{x}_{j}\right)  ^{2}~>0~\text{\ for}\ \ ij=12,34~.\nonumber
\end{align}
The best intuition for the above configuration comes from thinking of the CFT
positions $\mathbf{x}_{i}$ as points on the boundary of global AdS, as in
figure \ref{fig8} in the introduction. The conditions (\ref{in100}) then
corresponds to a true Lorentzian scattering process in the dual AdS geometry.
We shall also require that%
\begin{equation}
\left(  \mathbf{x}_{i}-\mathbf{x}_{j}\right)  ^{2}~>0~\text{\ for}\ \ ij=13,24~.
\label{in100BIS}%
\end{equation}
This condition is not essential, but streamlines considerably our discussion \cite{Paper4}. 
For such configurations, shown in figure \ref{fig7}a, the relevant reduced amplitude is 
given by a specific analytic continuation $\hat{\mathcal{A}}$ of $\mathcal{A}$, as described 
in figure \ref{fig7}b and in detail in \cite{Paper3, Paper4}. We shall be
interested in the study of the amplitude $\hat{\mathcal{A}}\left(  z,\bar
{z}\right)  $ in the limit 
$\left(\mathbf{x}_{1}-\mathbf{x}_{3}\right)^{2},\left(\mathbf{x}_{2}-\mathbf{x}_{4}\right)^{2}\rightarrow0$, 
which
implies $z,\bar{z}\rightarrow0$. 

To clarify the underlying geometry of the Lorentzian amplitude, it is
best to introduce the concept of transverse conformal group \cite{Paper4}.
Consider the correlator (\ref{in000}) as a function of two points,
$\mathbf{x}_{3}$ and $\mathbf{x}_{2}$, fixing the positions of $\mathbf{x}%
_{1}$ and $\mathbf{x}_{4}$. The subgroup of the conformal group which leaves
the points $\mathbf{x}_{1}$ and $\mathbf{x}_{4}$ fixed is given by $SO\left(
1,1\right)  \times SO\left(  3,1\right)  $. This fact is manifest if we use
conformal symmetry to send the point $\mathbf{x}_{1}$ to the origin and the
point $\mathbf{x}_{4}$ to infinity, which can be achieved, for instance, by
first translating $\mathbf{x}_{i}\rightarrow\mathbf{x}_{i}-\mathbf{x}_{1}$ and
then by performing a special conformal transformation $\mathbf{y\rightarrow
}\left(  \mathbf{y\,a}^{2}-\mathbf{a\,y}^{2}\right)  /\left(  \mathbf{a-y}%
\right)  ^{2}$, with $\mathbf{y=x}_{i}-\mathbf{x}_{1}$ and $\mathbf{a\,=x}%
_{4}-\mathbf{x}_{1}$. Under these transformations
the points $\mathbf{x}_{3}$ and
$\mathbf{x}_{2}$ are mapped, respectively, to $-\mathbf{x}$ and $\mathbf{\bar
{x}/\bar{x}}^{2}$, with
\begin{align*}
\mathbf{x}  &  =\frac{\left(  \mathbf{x}_{4}-\mathbf{x}_{1}\right)  \left(
\mathbf{x}_{3}-\mathbf{x}_{1}\right)  ^{2}-\left(  \mathbf{x}_{3}%
-\mathbf{x}_{1}\right)  \left(  \mathbf{x}_{4}-\mathbf{x}_{1}\right)  ^{2}%
}{\left(  \mathbf{x}_{4}-\mathbf{x}_{3}\right)  ^{2}}~,\\
\mathbf{\bar{x}}  &  =\frac{\left(  \mathbf{x}_{2}-\mathbf{x}_{1}\right)
\left(  \mathbf{x}_{4}-\mathbf{x}_{1}\right)  ^{2}-\left(  \mathbf{x}%
_{4}-\mathbf{x}_{1}\right)  \left(  \mathbf{x}_{2}-\mathbf{x}_{1}\right)
^{2}}{\left(  \mathbf{x}_{4}-\mathbf{x}_{1}\right)  ^{2}\left(  \mathbf{x}%
_{2}-\mathbf{x}_{1}\right)  ^{2}}~.
\end{align*}
The vectors $\mathbf{x}$ and $\mathbf{\bar{x}}$ are defined up to the residual
conformal symmetry. This is given by $SO\left(  3,1\right)  $ rotations, under
which $\mathbf{x}$ and $\mathbf{\bar{x}}$ transform as vectors, and 
by $SO\left( 1,1\right)$ dilatations, under which 
$\mathbf{x\rightarrow}\lambda\mathbf{x}$ and
$\mathbf{\bar{x}\rightarrow\bar{x}}/\lambda$. Moreover, for the kinematics
(\ref{in100}) and (\ref{in100BIS}), $\mathbf{x}$ is in the future of 
$\mathbf{\bar{x}/\bar{x}}^{2}$ and
\[
\mathbf{x,\bar{x}}\in\mathrm{M~}\text{\textrm{,}}
\]
where $\mathrm{M}\in\mathbb{M}^{4}$ is the future Milne wedge. The reduced
amplitude can then be written as
\[
\hat{\mathcal{A}}\left(  \mathbf{x,\bar{x}}\right)
\]
and depends only on the $SO\left(  1,1\right)  \times SO\left(  3,1\right)  $
conformally invariant cross-ratios%
\[
z\bar{z}=\mathbf{x}^{2}\mathbf{\bar{x}}^{2}%
~,~\ \ \ \ \ \ \ \ \ \ \ \ \ \ \ \ z+\bar{z}=-2\mathbf{x\cdot\bar{x}}~.
\]

We are interested in the limit $\mathbf{x,\bar{x}}\rightarrow0$ of the reduced
amplitude. As shown in \cite{Paper2}, this limit is not dominated by the OPE and
therefore by operators of lowest conformal dimension, as in the Euclidean
version of the theory, but by the exchanged operators of maximal spin. Whenever
the exchanged spin is unbounded, the limiting $\mathbf{x,\bar{x}}\rightarrow0$
behavior must be analyzed using Regge techniques \cite{Paper4}. 
In the presence of a Regge
pole with trajectory $j\left(  \nu\right)  $, the limit of the reduced
amplitude reads\footnote{For simplicity of notation, here and in the rest of the paper we write 
$\left\vert 4 \mathbf{x} \mathbf{\bar{x}} \right\vert$ instead of the more cumbersome expression 
$ 4\left\vert\mathbf{x}\right\vert\left\vert\mathbf{\bar{x}}\right\vert$.} 
\[
\hat{\mathcal{A}}\left(  \mathbf{x},\mathbf{\bar{x}}\right)  \simeq2\pi i~\int
d\nu~\left(  -\right)  ^{j\left(  \nu\right)  }~\alpha\left(  \nu\right)
~\left\vert 4\mathbf{x\bar{x}}\right\vert ^{1-j\left(  \nu\right)  }%
~\Omega_{i\nu}\left(  \mathbf{x},\mathbf{\bar{x}}\right)  ~,
\]
where $\alpha\left(  \nu\right)  $ is the pole residue and where $\Omega
_{i\nu}\left(  \mathbf{x},\mathbf{\bar{x}}\right)  $, given explicitly in
\cite{Paper4} and in section \ref{secAmpPosSpace} of this paper, computes radial Fourier
transforms in the transverse hyperbolic space $\mathrm{H}_{3}\subset
\mathrm{M}$ and solves the homogeneous equation $\left(  \square
_{\mathrm{H}_{3}}+\nu^{2}+1\right)  \Omega_{i\nu}=0$. In CFT's with an
AdS$_{5}$ string dual, the hyperbolic space $\mathrm{H}_{3}$ plays the role of
the space transverse to the interaction and $\ell^{-2}\square_{\mathrm{H}_{3}%
}$ measures transverse momentum transfer, with $\ell$ the AdS radius.
Therefore, for large~$\ell$, we may think of $\nu/\ell$ as momentum transfer
in AdS$_{5}$.

\subsection{$\mathcal{N}=4$ super Yang Mills}

We shall focus our attention on the canonical example of $\mathcal{N}=4$,
$SU\left(N\right)$ SYM with 't Hooft coupling
$g^{2}=g_{\mathrm{YM}}^{2}N$. The theory is dual to IIB strings on
AdS$_{5}\times$S$_{5}$, with AdS radius $\ell=\sqrt{\alpha^{\prime}g}$ and
$5$--dimensional Newton constant $G=\pi\ell^{3}/2N^{2}$. In particular, as a
basic example, let us consider the correlator (\ref{in000}), with%
\[
\mathcal{O}_{1}=c~\mathrm{Tr}\left(  Z^{2}\right)
~\,,~\ \ \ \ \ \ \ \ \ \ \ \ \ \ \ \ \ \ \ \ \ \ \ \mathcal{O}_{2}%
=c~\mathrm{Tr}\left(  W^{2}\right)  ~,
\]
where $Z$ and $W$ are two of the three complex scalar fields of the theory and
$c$ is a normalization constant fixed so that the $2$--point functions
$\left\langle \mathcal{O}_{1}\mathcal{O}_{1}^{\star}\right\rangle $ and
$\left\langle \mathcal{O}_{2}\mathcal{O}_{2}^{\star}\right\rangle $ are
canonically normalized to $1/\left(  \mathbf{x}_{i}-\mathbf{x}_{j}\right)
^{4}$. The operators $\mathcal{O}_{i}$ are chiral primaries and are not
renormalized, with $\Delta_{1}=\Delta_{2}=2$. Therefore, the reduced amplitude
$\mathcal{A}$ will read
\[
\mathcal{A}=1+\frac{1}{N^{2}}\,\mathcal{A}_{\mathrm{planar}}+\cdots~,
\]
where $1$ represents the disconnected part $\left\langle \mathcal{O}%
_{1}\mathcal{O}_{1}^{\star}\right\rangle \left\langle \mathcal{O}%
_{2}\mathcal{O}_{2}^{\star}\right\rangle $, whereas $\mathcal{A}%
_{\mathrm{planar}}$ represents the planar contribution to the amplitude, dual
to tree--level string interactions. The planar contribution depends
non--trivially on the 't Hooft coupling $g^{2}$ and should be dominated, in
the $z,\bar{z}\rightarrow0$ Lorentzian regime described above, by the Regge
pole associated to the exchange of the tower of massive string states of
lowest twist. In particular, we expect that the planar contribution should be
given by%
\begin{equation}
\hat{\mathcal{A}}_{\mathrm{planar}}\left(  \mathbf{x},\mathbf{\bar{x}}\right)
\simeq2\pi i~\int d\nu~\left(  -\right)  ^{j\left(  \nu,g\right)  }%
~\alpha\left(  \nu,g\right)  ~\left\vert 4\mathbf{x\bar{x}}\right\vert
^{1-j\left(  \nu,g\right)  }~\Omega_{i\nu}\left(  \mathbf{x},\mathbf{\bar{x}%
}\right)  ~, \label{b100}%
\end{equation}
where we have explicitly shown the $g$ dependence of the trajectory $j\left(
\nu,g\right)  $ and of the residue function $\alpha\left(  \nu,g\right)  $.

\subsection{Large 't Hooft coupling}

At large 't Hooft coupling, the dominant Regge trajectory is dual to graviton
exchange in AdS. As shown in \cite{PS1, Paper4}, the trajectory has a large
$g$ expansion given by%
\[
j\left(  \nu,g\right)  =2-\frac{4+\nu^{2}}{2g}+\cdots~.
\]
Moreover, in the limit $g\rightarrow\infty$, the residue function
$\alpha\left(  \nu,g\right)  $ is given by \cite{Paper4}
\[
\alpha\left(  \nu,g\right)  \simeq-\pi~V_{\mathrm{\min}}\left(  \nu,j=2\right)
\frac{1}{4+\nu^{2}}\,\bar{V}_{\mathrm{\min}}\left(  \nu,j=2\right)
~,\ \ \ \ \ \ \ \ \left(  g\rightarrow\infty\right)\ .
\]
The term $1/\left(  4+\nu^{2}\right)  $ represents the graviton propagator,
dual to the CFT stress--energy tensor of dimension $4$, which
corresponds\footnote{As discussed in \cite{Paper4}, we normalize $\nu\,$\ so
that CFT dimensions are given by $2+i\nu$.} to $\nu=-2i$. The function
$V_{\min}=\bar{V}_{\min}$ is given explicitly by
\[
V_{\min}\left(  \nu,j\right)  =4^{j-1}~\Gamma\left(  1+\frac{j+i\nu}%
{2}\right)  \Gamma\left(  1+\frac{j-i\nu}{2}\right)  ~
\]
and represents the minimal coupling of the dimension $2$ external scalars to
the exchanged trajectory of spin $j=j\left(  \nu,g\right)  $. In the limit
$g\rightarrow\infty$, one has that $j\rightarrow2$ corresponding to the usual
gravitational field, so that%
\[
\alpha\left(  \nu,g\right)  \simeq-\frac{\pi^{3}}{4}\frac{\nu^{2}\left(
4+\nu^{2}\right)  }{\sinh^{2}\left(  \frac{\pi\nu}{2}\right)  }%
~,\ \ \ \ \ \ \ \ \ \ \ \ \ \ \ \ \ \ \ \left(  g\rightarrow
\infty\right)\ .
\]

\subsection{Weak 't Hooft coupling}

The main focus of this paper is devoted, though, to the analysis of
$\mathcal{A}_{\mathrm{planar}}$ at weak coupling $g\rightarrow0$. The planar
amplitude $\mathcal{A}_{\mathrm{planar}}$ has been computed to order $g^{4}$
in \cite{FourPT}, with explicit result%
\begin{align}
&  \mathcal{A}_{\mathrm{planar}}\left(  z,\bar{z}\right)  =
-\frac{g^{2}}{2\pi^{2}}\,\Phi_{1}\left(  z,\bar{z}\right)  +\frac{g^{4}}{16\pi^{4}}%
\frac{2+2z\bar{z}-z-\bar{z}}{4z\bar{z}}\,\Phi_{1}^{2}(z,\bar{z})\label{in600}\\
&  +\frac{g^{4}}{16\pi^{4}}\frac{z\bar{z}}{z-\bar{z}}\left[  \Phi_{2}%
(z,\bar{z})-\Phi_{2}(1-z,1-\bar{z})-\Phi_{2}\left(  \frac{z}{z-1},\frac
{\bar{z}}{\bar{z}-1}\right)  \right]  ~,\nonumber
\end{align}
where%
\begin{align*}
\Phi_{1}(z,\bar{z})  &  =\frac{z\bar{z}}{z-\bar{z}}\left[  2\mathrm{Li_{2}%
}(z)-2\mathrm{Li_{2}}(\bar{z})+\log(z\bar{z})\log\frac{1-z}{1-\bar{z}}\right]
~,\\
\Phi_{2}(z,\bar{z})  &  =6\Big[  \mathrm{Li_{4}}(z)-\mathrm{Li_{4}}(\bar
{z})\Big]  -3\log z\bar{z}~\Big[  \mathrm{Li_{3}}(z)-\mathrm{Li_{3}}%
(\bar{z})\Big]  +\\
&  +\frac{1}{2}\log^{2}z\bar{z}~\Big[  \mathrm{Li_{2}}(z)-\mathrm{Li_{2}%
}(\bar{z})\Big]  ~.
\end{align*}
Using the fact that for  $z\rightarrow0$ the analytic
continuation of $\widehat{\mathrm{Li}}\mathrm{_{n}}(z)$ is given by
$\widehat{\mathrm{Li}}\mathrm{_{n}}(z)\simeq-2\pi
i~\ln^{n-1}\left(  -z\right)  /\left(  n-1\right)  !$, it is easy to show that,
in this limit, the Lorentzian
amplitude $\hat{\mathcal{A}}_{\mathrm{planar}}$ is dominated by the term
proportional to $g^{4}\Phi_{1}^{2}/z\bar{z}$, and it is explicitly given by
\begin{equation}
\hat{\mathcal{A}}_{\mathrm{planar}}\simeq-\frac{g^{4}}{8\pi^{2}}\frac{z\bar
{z}}{(z-\bar{z})^{2}}\log^{2}\frac{\bar{z}}{z}
~,~\ \ \ \ \ \ \ \ \ \ \ \left(  z,\bar{z}\rightarrow0\right)\ .
\label{s2}
\end{equation}
The above expression is invariant under rescalings $z,\bar{z}\rightarrow
\lambda z,\lambda\bar{z}$, and it therefore corresponds to the contribution of
a leading Regge pole of spin $j=1$. Moreover, explicitly computing the radial
Fourier transform in the transverse space $\mathrm{H}_{3}$, one may show that
(\ref{s2}) corresponds to\footnote{Defining $\bar{z}/z=e^{-2\rho}$, one has
that $\hat{\mathcal{A}}_{\mathrm{planar}}=-g^{4}\rho^{2}/\left(  8\pi^{2}%
\sinh^{2}\rho\right)  $. Since $\hat{\mathcal{A}}_{\mathrm{planar}}$ is given
by (\ref{b100}) for $j=1$, one has that $\alpha=\frac{2i}{\nu}\int_{0}%
^{\infty}d\rho\sin\nu\rho~\sinh\rho~\hat{\mathcal{A}}_{\mathrm{planar}}$, as
shown in \cite{Paper4}.}
\begin{equation}
\alpha\left(  \nu,g\right)  \simeq-\frac{i}{4\pi}~V\left(  \nu\right)
\frac{\tanh\frac{\pi\nu}{2}}{\nu}\,\bar{V}\left(  \nu\right)
~,\ \ \ \ \ \ \ \ \ \ \ \ \ \left(  g\rightarrow0\right)\ ,
\label{in500}
\end{equation}
where
\begin{equation}
V\left(  \nu\right)  =\bar{V}\left(  \nu\right)  =\frac{\pi g^{2}}{2}\frac
{1}{\cosh\frac{\pi\nu}{2}}~~. \label{in501}%
\end{equation}
As we shall review in more detail in section \ref{SecBFKL}, the above result
is dominated by the Regge pole of the perturbative hard BFKL Pomeron
\cite{BFKL, Lipatov, LipatovRev}, with trajectory given by the famous
expression
\[
j\left(  \nu,g\right)  =1+\frac{g^{2}}{4\pi^{2}}\left(  2\Psi\left(  1\right)
-\Psi\left(  \frac{1+i\nu}{2}\right)  -\Psi\left(  \frac{1-i\nu}{2}\right)
\right)  +\cdots~,
\]
which converges to $j=1$ for $g\rightarrow0$. In the next section, we shall
formulate the usual BFKL formalism completely in position space and
explicitly derive (\ref{in500}). The factor $\tanh\left(  \pi\nu/2\right)
/\nu$ corresponds to the pomeron propagator, whereas $V\left(  \nu\right)  $
and $\bar{V}\left(  \nu\right)  $ correspond to the couplings of external states
to the pomeron, usually called impact factors in the literature. 
We shall derive the explicit
leading order expressions (\ref{in501}) directly in perturbation theory in
position space in section \ref{SecN4}, thus rederiving (\ref{in500}) without
the need of the full result (\ref{in600}).

The use of position space techniques streamlines considerably the usual
computations based on the momentum space BFKL impact factors. In particular, we
shall show how the position space formalism, which uses heavily the invariance
under the transverse conformal group $SO\left(  1,1\right)  \times SO\left(
3,1\right)  $, immediately implies that only the $n=0$ part of the
BFKL\ kernel gives non--vanishing overlap with impact factors of scalar
external states.

\subsection{Eikonalization of the pomeron exchange and saturation}

Let us conclude this introductory section with some more speculative
considerations. Recall from \cite{Paper4} that the contribution from a single
pomeron exchange grows too fast at high energy and eventually violates the
unitarity bounds. At large impact parameters, we expect that one should be
able to restore unitarity by considering multiple pomeron exchanges using
eikonal methods. The CFT extension of the usual eikonal resummation, which
corresponds dually to eikonalization in the dual AdS geometry, was developed
in \cite{Paper2, Paper3, PS2} and was generalized to Regge pole exchanges in
\cite{Paper4, PS3}. Let us first recall the basic facts. 
In the regime of small $\mathbf{x},\mathbf{\bar{x}}$, the CFT
amplitude $\hat{\mathcal{A}}\left(  \mathbf{x},\mathbf{\bar{x}}\right)$
admits an impact parameter representation in AdS given by \cite{Paper2,
Paper4}
\begin{equation}
\hat{\mathcal{A}}\left(  \mathbf{x},\mathbf{\bar{x}}\right)  =
\frac{4\left\vert \mathbf{x\bar{x}}\right\vert ^{4}}{\pi^{2}}\int_{\mathrm{M}%
}d\mathbf{y}d\mathbf{\bar{y}}~e^{-2i\mathbf{x\cdot y-}2i\mathbf{\bar{x}%
\cdot\bar{y}}}~e^{-2\pi i~\Gamma\left(  \mathbf{y},\mathbf{\bar{y}}\right)
}~, \label{s3}%
\end{equation}
where the Fourier integral $d\mathbf{y}d\mathbf{\bar{y}}$ is supported only in
the future Milne cone $\mathrm{M}$. The function $\Gamma\left(  \mathbf{y}%
,\mathbf{\bar{y}}\right)  $ plays the role of the phase shift and depends on
the $SO\left(  1,1\right)  \times SO\left(  3,1\right)  $ invariants
$s=\left\vert 4\mathbf{y\bar{y}}\right\vert $ and $\cosh r=-\mathbf{y\cdot
\bar{y}/}\left\vert \mathbf{y\bar{y}}\right\vert $, which correspond to energy--squared
and impact parameter in the dual AdS geometry. As in flat space scattering,
the impact parameter representation approximates the AdS (conformal) partial
wave decomposition for large values of the impact parameter and energy. In
analogy with flat space, AdS unitarity should be diagonalized by the partial
wave decomposition and should simply corresponds to the requirement
$\operatorname{Im}\Gamma\left(  \mathbf{y},\mathbf{\bar{y}}\right)  \leq0$.
Let us note, though, that the status of unitarity in AdS interactions is not
on the same firm theoretical grounds as the corresponding statements in flat
space, due to the lack of asymptotic states and of an explicitly unitary $S$--matrix.

When $\Gamma=0$, there is no AdS interaction and $\hat{\mathcal{A}}=1$. We may
then define the planar phase shift $\Gamma_{\mathrm{planar}}$  by%
\[
\frac{1}{N^{2}}\,\hat{\mathcal{A}}_{\mathrm{planar}}\left(  \mathbf{x}%
,\mathbf{\bar{x}}\right)  = -2\pi i~\frac{4\left\vert \mathbf{x\bar{x}%
}\right\vert ^{4}}{\pi^{2}}\int_{\mathrm{M}}d\mathbf{y}d\mathbf{\bar{y}%
}~e^{-2i\mathbf{x\cdot y-}2i\mathbf{\bar{x}\cdot\bar{y}}}~\Gamma
_{\mathrm{planar}}\left(  \mathbf{y},\mathbf{\bar{y}}\right)  ~.
\]
Whenever the planar amplitude is dominated, for small $\mathbf{x},\mathbf{\bar{x}}$, by a Regge pole and is 
given by (\ref{b100}), the above expression can be inverted to get
\[
\Gamma_{\mathrm{planar}}\left(  \mathbf{y},\mathbf{\bar{y}}\right)
\simeq\frac{1}{N^{2}}\int d\nu~\beta\left(  \nu,g\right)  ~\left\vert
4\mathbf{y\bar{y}}\right\vert ^{~j\left(  \nu,g\right)  -1}~\Omega_{i\nu
}\left(  \mathbf{y},\mathbf{\bar{y}}\right)  ~,
\]
valid for large $\mathbf{y},\mathbf{\bar{y}}$, with $\beta( \nu,g)$ defined by
\[
\alpha\left(  \nu,g\right)  =~V_{\mathrm{\min}}\big(  \nu,j\left(
\nu,g\right)  \big)  ~\beta\left(  \nu,g\right)  ~\bar{V}_{\mathrm{\min}}
\big(  \nu,j\left(  \nu,g\right)  \big)  ~.\ \
\]
Note that we have
\begin{align}
\beta\left(  \nu,g\right)   &  \simeq-\frac{\pi}{4+\nu^{2}}%
~,\ \ \ \ \ \ \ \ \ \ \ \ \ \ \ \ \ \ \ \ \ \ \ \ \ \ \ \ \ \ \ \left(
g\rightarrow\infty\right)\ , \nonumber\\
\beta\left(  \nu,g\right)   &  \simeq-\frac{ig^{4}}{\pi}\frac{\tanh\frac
{\pi\nu}{2}}{\nu\left(  1+\nu^{2}\right)  ^{2}}%
~,\ \ \ \ \ \ \ \ \ \ \ \ \ \  \ \ \ \ \ \ \ \left(  g\rightarrow0\right)\ .
\label{eq3001}
\end{align}
In the eikonal approximation, valid in principle for large values of the AdS
energy--squared $s$ and impact parameter $r$, the full phase shift $\Gamma$ is
approximated by the planar contribution $\Gamma_{\mathrm{planar}}$. The
eikonal amplitude, resumming multi--pomeron exchanges, may then be written as
\begin{equation}
\hat{\mathcal{A}}_{\mathrm{eikonal}}\left(  \mathbf{x},\mathbf{\bar{x}%
}\right)  \simeq\frac{4\left\vert \mathbf{x\bar{x}}\right\vert ^{4}}{\pi^{2}%
}\int_{\mathrm{M}}d\mathbf{y}d\mathbf{\bar{y}}~e^{-2i\mathbf{x\cdot
y-}2i\mathbf{\bar{x}\cdot\bar{y}}}~e^{-2\pi i~\Gamma_{\mathrm{planar}}\left(
\mathbf{y},\mathbf{\bar{y}}\right)  }~. \label{eq3000}%
\end{equation}
The eikonal expression (\ref{eq3000}) would then automatically implement
unitarity both at weak and at strong coupling for $\operatorname{Im}%
\Gamma_{\mathrm{planar}}\leq0$. In particular, the $g\rightarrow 0$ limit
(\ref{eq3001}) given by
\[
\Gamma_{\mathrm{planar}}\left(  \mathbf{y},\mathbf{\bar{y}}\right)
\simeq-\frac{ig^{4}}{\pi N^{2}}\int d\nu~\frac{\tanh\frac{\pi\nu}{2}}%
{\nu\left(  1+\nu^{2}\right)  ^{2}}~\Omega_{i\nu}\left(  \mathbf{y}%
,\mathbf{\bar{y}}\right)  ,\ \ \ \ \ \ \ \ \ \left(
g\rightarrow0\right)\ ,
\]
has negative imaginary part, as expected.

An important unresolved issue concerns the relation of the $5$--dimensional
AdS eikonal expression (\ref{eq3000}) to the standard eikonalization of the
correlator (\ref{in000}) in four dimensions. In fact, it is well known that
unitarization of the weak coupling BFKL pomeron exchange using $4$--dimensional 
eikonal techniques fails to reproduce the correct physics at large
energies and must be supplemented by the far more complex analysis of
non--linear pomeron interactions, which in turn lead to the phenomenon of
gluon saturation in the structure functions of the scattering states (see
\cite{Iancu} for reviews and for an extensive list of relevant references). On
the other hand, multi--pomeron interactions have never been analyzed using the
AdS expression (\ref{eq3000}). It is quite reasonable that, for a certain
range of AdS impact parameters, the planar approximation to the phase shift is
still valid even when the phase shift is of order one and a single exchange
violates the unitarity bound. Let us note that eikonal resummations are
possibly the simplest technique to analyze the ambient geometry, since it is
inherently based on geodesic motion in the spacetime where interactions take
place. Moreover, saturation effects, where non--linear pomeron interactions
are relevant, have already been seen at present accelerators. It is then quite
conceivable that, for carefully chosen external kinematics, interactions are
approximated by expression (\ref{eq3000}), thus showing experimentally
the duality between field theories and gravity. We plan to
address some of these issues in \cite{Paper6}.

\section{BFKL Analysis in Position Space\label{SecBFKL}}

\subsection{The BFKL kernel at vanishing coupling\label{s100}}

\begin{figure}[ptb]
\begin{center}
\includegraphics[height= 1.3in]{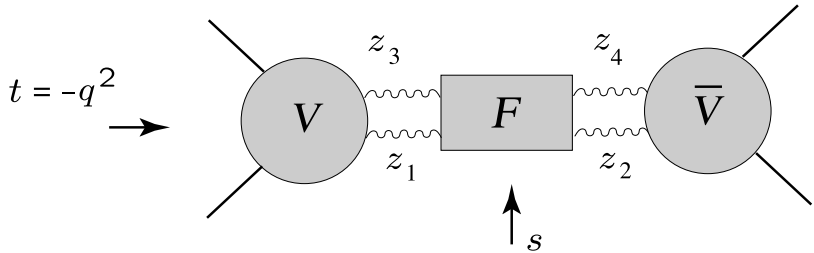}
\end{center}
\caption{Exchange of a BFKL pomeron. At leading order in the coupling
constant, the kernel $F$ is given by the exchange of a pair of transverse
gluons in a color singlet state.}%
\label{fig2}%
\end{figure}

High energy interactions in gauge theories are dominated by hard pomeron
exchange for $s\gg\left\vert t\right\vert \gg\Lambda_{\text{QCD}}$. In the
Born approximation, the leading contribution at high energies comes from the
exchange of a pair of gluons in a color singlet state, and
amplitudes\footnote{In order to have a consistent notation when discussing CFT
correlators, we incorporate, in the definition of the amplitude, the factor of
$i$, using the convention $1+\mathcal{A}$ as opposed to the more standard
field theory convention $1+i\mathcal{A}$.} are conveniently written as
\cite{BFKL,LipatovRev}
\begin{equation}
-s~\int_{\mathbb{E}^{2}}dz_{1}\cdots dz_{4}~V_{q}\left(  z_{1},z_{3}\right)
\;F\left(  z_{1},z_{3},z_{2},z_{4}\right)  \;\bar{V}_{q}\left(
z_{2},z_{4}\right)  ~. \label{e100}%
\end{equation}
Let us describe qualitatively the main features of (\ref{e100}),
using as an aid figure \ref{fig2}. First of all, the overall energy dependence
$s$ shows that we are exchanging a Regge pole with effective spin $1$. At high
energies, the exchanged gluons are essentially transverse, and are replaced by
a pair of massless propagators
\begin{equation}
F\left(  z_{1},z_{3},z_{2},z_{4}\right)  =2\ln\left(  z_{1}-z_{2}\right)
^{2}\ln\left(  z_{3}-z_{4}\right)  ^{2}~ \label{e300}%
\end{equation}
in transverse space $\mathbb{E}^{2}$, where the $z_{i}\in\mathbb{E}^{2}$ are
the gluon transverse positions. The coupling of the pair of gluons to the
scattering states is, on the other hand, described by the impact
factors $V_{q}\left(  z_{1},z_{3}\right)  $ and $\bar{V}_{q}\left(
z_{2},z_{4}\right)  $. These factors depend on the transverse momentum
transfer $q$ in $\mathbb{E}^{2}$ and also on other features of the external
incoming and outgoing states, like virtualities and polarizations, which we do
not show explicitly. Whenever the scattering states have vanishing color
charge, the impact factors satisfy the infrared finiteness condition
\cite{LipatovRev}
\begin{equation}
\int_{\mathbb{E}^{2}}~dz_{1}~V_{q}\left(  z_{1},z_{3}\right)  =\int
_{\mathbb{E}^{2}}~dz_{3}~V_{q}\left(  z_{1},z_{3}\right)  =0~, \label{e200}%
\end{equation}
and similarly for $\bar{V}_{q}\left(  z_{2},z_{4}\right)  $. Finally, note
that the amplitude is mostly real (imaginary in the usual field theory
convention), leading to an imaginary phase shift.

Let us first concentrate on the pomeron kernel (\ref{e300}) describing the
propagation of the two transverse gluons. At finite 't Hooft coupling
$g^{2}=g_{YM}^{2}N$, the leading corrections to (\ref{e300}) are described by
the BFKL equation \cite{BFKL,LipatovRev}. As noted by Lipatov in
\cite{Lipatov}, the BFKL equation is invariant under transverse conformal
transformations $SO\left(  3,1\right)  $ of $\mathbb{E}^{2}$ if we assume that
$F$ transforms like a $4$--point function of scalar primaries of vanishing
dimension. It is then natural to look for solutions depending on the
transverse harmonic cross--ratios
\begin{equation}
\frac{\left(  z_{1}-z_{3}\right)  ^{2}\left(  z_{2}-z_{4}\right)  ^{2}%
}{\left(  z_{1}-z_{2}\right)  ^{2}\left(  z_{3}-z_{4}\right)  ^{2}%
}~,~\ \ \ \ \ \ \ \ \ \ \ \ \ \ \frac{\left(  z_{1}%
-z_{4}\right)  ^{2}\left(  z_{2}-z_{3}\right)  ^{2}}{\left(  z_{1}%
-z_{2}\right)  ^{2}\left(  z_{3}-z_{4}\right)  ^{2}}~. \label{e1000}%
\end{equation}
Clearly, (\ref{e300}) is not invariant under conformal transformations of
$\mathbb{E}^{2}$. However, we are free to add to the BFKL kernel any
function which is independent of at least one of the $z_{i}$'s, since physical
amplitudes (\ref{e100}) are obtained by integrating against impact factors
satisfying (\ref{e200}). Therefore, we may substitute the kernel (\ref{e300})
with the equivalent conformally invariant function
\begin{equation}
F~\mathbf{=}\,\ln\frac{\left(  z_{1}-z_{3}\right)  ^{2}\left(  z_{2}%
-z_{4}\right)  ^{2}}{\left(  z_{1}-z_{2}\right)  ^{2}\left(  z_{3}%
-z_{4}\right)  ^{2}}\,\ln\frac{\left(  z_{1}-z_{4}\right)  ^{2}\left(
z_{2}-z_{3}\right)  ^{2}}{\left(  z_{1}-z_{2}\right)  ^{2}\left(  z_{3}%
-z_{4}\right)  ^{2}}~. \label{e1200}%
\end{equation}

\subsection{Explicit transverse conformal invariance\label{SecConfInv}}

In order to render the transverse conformal invariance
manifest, it is best to work in Minkowski space $\mathbb{M}^{4}$ on which 
the transverse conformal group $SO(3,1)$, introduced in section \ref{SecGeneral}, 
acts naturally.
This discussion entirely parallels the case of the conformal group $SO(d,2)$ 
of $d$--dimensional Minkowski spacetime, whose action on the light--cone 
of an embedding $\mathbb{E}^{d,2}$ space is linear,
as reviewed in  \cite{Paper1,Paper2}.

\begin{figure}[ptb]
\begin{center}
\includegraphics[height= 1.5in]{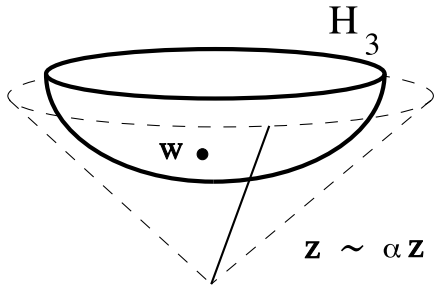}
\end{center}
\caption{Hyperbolic $3$--space H$_3$ seen as the unit mass--shell in
$\mathbb{M}^4$, given by points $\mathbf{w}\in\mathrm{M}$ with $\mathbf{w}^{2}=-1$.
The boundary $\partial\mathrm{H}_3$ is then naturally identified with lines in the
light--cone, given by points $\mathbf{z}\in\partial\mathrm{M}$ with $\mathbf{z}^{2}=0$, 
defined up to rescalings $\mathbf{z}\sim\alpha\mathbf{z}$.}%
\label{fig10}%
\end{figure}

Let us recall some basic
notation schematica, denote with $\mathrm{M}\subset\mathbb{M}^{4}$ the future Milne
wedge, with $\partial\mathrm{M}\subset\mathbb{M}^{4}$ the future light--cone
and with $\mathrm{H}_{3}\subset\mathrm{M}\subset\mathbb{M}^{4}$ the hyperbolic
$3$--space of points $\mathbf{w}\in\mathrm{M}$ with $\mathbf{w}^{2}=-1$,
holographically dual to the transverse space of the gauge theory. 
The boundary of $\mathrm{H}_{3}$ can also be described 
invariantly by using the embedding
space $\mathbb{M}^{4}$. More precisely\footnote{In \cite{Paper1}, the
analogous statements where discussed for the boundary of the full AdS$_{5}$,
seen as light rays in the embedding space $\mathbb{E}^{4,2}$.}, 
we may think of transverse space as light-rays in $\partial\mathrm{M}$,
i.e. points $\mathbf{z}=\left(  z^{+},z^{-},z\right)  \in\mathbb{M}^{4}$ such
that $\mathbf{z}^{2}=0$ and $z^{\pm}>0$, identifying points $\mathbf{z}%
\sim\alpha\mathbf{z}$ related by a positive rescaling factor $\alpha$,
as represented in figure \ref{fig10}.
Transverse space is then recovered by\ taking an arbitrary slice of the
light--cone $\partial\mathrm{M}$, choosing a specific representative for each
ray (for an extensive discussion of this point, see for instance \cite{PeneThesis}).
We shall denote with $\partial\mathrm{H}_{3}$ any given choice of such slice.
The standard space $\mathbb{E}^{2}$ is recovered with the usual Poincar\'{e}
choice $z^{+}=1$, so that a generic point is parameterized by points
$z\in\mathbb{E}^{2}$ as
\begin{equation}
\mathbf{z}=\left(  1,z^{2},z\right)  ~. \label{e400}%
\end{equation}
Note that, for two points $\mathbf{z}_{i}$ and $\mathbf{z}_{j}$ of the form above,
the inner product
\[
\mathbf{z}_{ij}\equiv-2\mathbf{z}_{i}\cdot\mathbf{z}_{j}%
\]
computes the usual Euclidean distance $\left(  z_{i}-z_{j}\right)  ^{2}$, so
that the cross--ratios (\ref{e1000}) can be written invariantly as%
\[
\frac{\mathbf{z}_{13}\,\mathbf{z}_{24}}{\mathbf{z}_{12}\,\mathbf{z}_{34}%
},~\ \ \ \ \ \ \ \ \ \ \ \ \ \ \ \ \ \ \ \frac{\mathbf{z}%
_{14}\,\mathbf{z}_{23}}{\mathbf{z}_{12}\,\mathbf{z}_{34}}~,
\]
and the BFKL\ kernel becomes a function $F\left(  \mathbf{z}_{1}%
,\mathbf{z}_{2},\mathbf{z}_{3},\mathbf{z}_{4}\right)  $ of the $SO\left(
3,1\right)  $ invariants $\mathbf{z}_{i}\cdot\mathbf{z}_{j}~\left(  i\neq
j\right)  $, invariant under rescalings $\mathbf{z}_{i}\rightarrow\alpha
_{i}\mathbf{z}_{i}$.

More generally, consider a generic function $f\left(  \mathbf{w}_{1}%
,\cdots,\mathbf{w}_{n}\right)  $ invariant under $SO\left(  3,1\right)  $. It
will generically depend on the $n\left(  n+1\right)  /2$ invariants
$\mathbf{w}_{i}\cdot\mathbf{w}_{j}$. On the other hand, if we assume that $f$
has weight $\Delta_{i}$ in the $i$--th entry, scaling as%
\[
f\left(  \cdots,\alpha\mathbf{w}_{i},\cdots\right)  =\alpha^{-\Delta_{i}%
}f\left(  \cdots,\mathbf{w}_{i},\cdots\right)  ~,
\]
the number of independent invariants is reduced to $n\left(  n-1\right)  /2$
cross--ratios. Finally, if $m$ of the points $\mathbf{w}_{i}$ are boundary
points on the light-cone $\partial\mathrm{M}$ and therefore satisfy
$\mathbf{w}^{2}=0$, the total number of cross--ratios is reduced to%
\begin{equation}
\frac{1}{2}\,n\left(  n-1\right)  -m~. \label{e1100}%
\end{equation}
The BFKL\ kernel has $n=4$, $m=4$ and therefore it has $2$ independent cross--ratios, as any
CFT $4$--point correlator.

We may obtain conformally invariant functions via integration. More precisely,
we may consider the integral%
\[
\int_{\mathrm{H}_{3}}~d\mathbf{w}_{n}~f\left(  \cdots,\mathbf{w}_{n}\right)
\]
over hyperbolic space $\mathrm{H}_{3}$, which clearly defines a conformal
function of the remaining points $\mathbf{w}_{1},\cdots,\mathbf{w}_{n-1}$.
More subtle is to construct conformally invariant functions via integration 
over the boundary $\partial\mathrm{H}_{3}$, due to
the arbitrariness in the choice of slice of $\partial\mathrm{M}$. One can 
easily show that the integral
\[
\int_{\partial\mathrm{H}_{3}}~d\mathbf{w}_{n}~f\left(  \cdots,\mathbf{w}%
_{n}\right)
\]
is independent of the choice of slice, and therefore conformally invariant,
whenever $\Delta_{n}=2$, i.e. whenever
\[
f\left(  \cdots,\alpha\mathbf{w}_{n}\right)  =\alpha^{-2}f\left(
\cdots,\mathbf{w}_{n}\right)  ~.
\]

\subsection{The $n=0$ component of the BFKL propagator}

\begin{figure}[ht]
\begin{center}
\includegraphics[height=1.7757in]{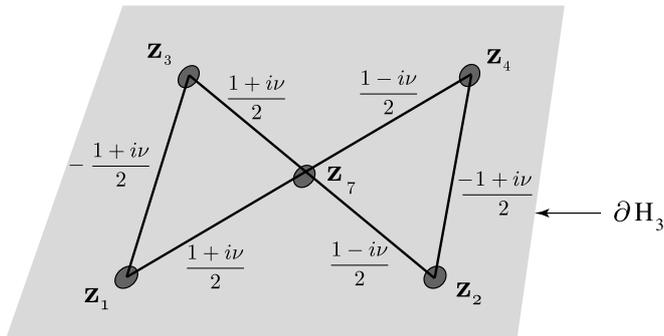}
\end{center}
\caption{Integral representation of the $n=0$ component of the BFKL kernel.}%
\label{fig3}%
\end{figure}

To analyze the $4$--point kernel (\ref{e1200}), it is best to construct
more basic conformal building blocks. Consider a conformal
function dependent on three boundary points $\mathbf{z}_{1},\mathbf{z}%
_{3},\mathbf{z}_{7}$, respectively with weights $0,0,1+i\nu$. There are no cross--ratios
and, up to a multiplicative constant, it is given uniquely by the $3$--point coupling of
scalar primaries%
\[
\left(  \frac{\mathbf{z}_{13}}{\mathbf{z}_{17}\mathbf{z}_{37}}\right)
^{\frac{1+i\nu}{2}}~.
\]
We may then consider the conformally invariant integral%
\begin{equation}
\int_{\partial\mathrm{H}_{3}}d\mathbf{z}_{7}~\left(  \frac{\mathbf{z}_{13}%
}{\mathbf{z}_{17}\mathbf{z}_{37}}\right)  ^{\frac{1+i\nu}{2}}\left(
\frac{\mathbf{z}_{24}}{\mathbf{z}_{27}\mathbf{z}_{47}}\right)  ^{\frac{1-i\nu
}{2}}~ \label{e1300}%
\end{equation}
shown in figure \ref{fig3}, where the total weight of the integrand in
$\mathbf{z}_{7}$ is correctly chosen to be $2$. For any value of $\nu$, the above
integral defines a conformal function of the four points points 
$\mathbf{z}_{1},\mathbf{z}_{2},\mathbf{z}_{3},\mathbf{z}_{4}$ with vanishing weights.

Consider now the leading BFKL propagator (\ref{e1200}). In general, it can be
written as a superposition of integrals of the form (\ref{e1300}) with a more
general integrand \cite{Lipatov}. The integrand itself is always constructed
from the product of $3$--point functions with an intermediate state, at the
point $\mathbf{z}_{7}$, of general spin $n\geq0$. However, as we
shall demonstrate later, whenever we compute amplitudes (\ref{e100}) with
external scalar operators, the contributions from the terms with $n>0$ vanish
due to conservation of transverse spin. The relevant $n=0$ part of the
BFKL\ kernel can then be written as a superposition of integrals of the form
(\ref{e1300}) with varying $\nu$. More precisely, we may replace (\ref{e1200})
with the expression \cite{Lipatov}%
\begin{equation}
\frac{4}{\pi^{2}}\int d\nu~\frac{\nu^{2}}{\left(  1+\nu^{2}\right)  ^{2}%
}\ \int_{\partial\mathrm{H}_{3}}d\mathbf{z}_{7}~\left(  \frac
{\mathbf{z}_{13}}{\mathbf{z}_{17}\mathbf{z}_{37}}\right)  ^{\frac{1+i\nu}{2}%
}\left(  \frac{\mathbf{z}_{24}}{\mathbf{z}_{27}\mathbf{z}_{47}}\right)
^{\frac{1-i\nu}{2}}~. \label{e3300}%
\end{equation}

\subsection{The amplitude in position space\label{secAmpPosSpace}}

\begin{figure}[h]
\begin{center}
\includegraphics[height=1.3005in]{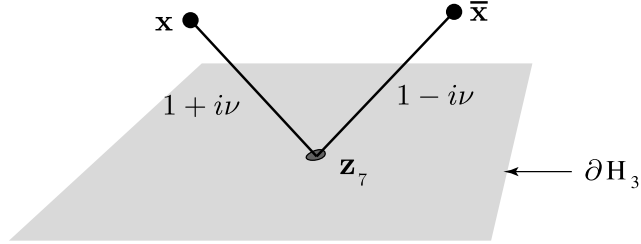}
\end{center}
\caption{Integral representation of the radial Fourier functions $\Omega
_{i\nu}\left(  \mathbf{x},\mathbf{\bar{x}}\right)  $ and of the basic planar
amplitude $\hat{\mathcal{A}}_{\mathrm{planar}}\left(  \mathbf{x}%
,\mathbf{\bar{x}}\right)  $.}%
\label{fig1}%
\end{figure}

In this paper, we are mostly interested in the analysis of the Lorentzian
amplitude%
\begin{equation}
\frac{1}{N^{2}}\,\hat{\mathcal{A}}_{\mathrm{planar}}\left(  \mathbf{x}%
,\mathbf{\bar{x}}\right)  \label{eq1111}%
\end{equation}
in position space, where we recall that $\mathbf{x}$ and $\mathbf{\bar{x}}$
are in the future Milne wedge $\mathrm{M}\subset\mathbb{M}^{4}$. 
In particular, we  focus our attention on the
$\mathbf{x},\mathbf{\bar{x}}\rightarrow0$ limit. As discussed in \cite{Paper4}
and reviewed in section \ref{SecG}, the limit of the planar amplitude is
dominated by a leading even signature Regge pole, whose spin $j\left(
\nu,g\right)  $ depends on the 't Hooft coupling $g^{2}$. For large $g$, the
pole corresponds to a reggeized spin--$2$ graviton exchanged in the bulk of
AdS, whereas for small $g$ the pole corresponds to the exchange of a hard BFKL
perturbative pomeron of spin approximately $1$. Therefore, recalling that the
reduced amplitude $\hat{\mathcal{A}}_{\mathrm{planar}}\left(  \mathbf{x}%
,\mathbf{\bar{x}}\right)  $ scales as $\left\vert \mathbf{x\bar{x}}\right\vert
^{1-j}$ for a pure spin--$j$ pole \cite{Paper4}, we deduce that, in the limit
$g\rightarrow0$, the amplitude $\hat{\mathcal{A}}_{\mathrm{planar}}\left(
\mathbf{x},\mathbf{\bar{x}}\right)  $ will be invariant under rescalings of
$\mathbf{x}$ and $\mathbf{\bar{x}}$ so that
\begin{equation}
\hat{\mathcal{A}}_{\mathrm{planar}}\left(  \alpha\mathbf{x},\mathbf{\bar{x}%
}\right)  =\hat{\mathcal{A}}_{\mathrm{planar}}\left(  \mathbf{x}%
,\alpha\mathbf{\bar{x}}\right)  =\hat{\mathcal{A}}_{\mathrm{planar}}\left(
\mathbf{x},\mathbf{\bar{x}}\right)
~,~\ \ \ \ \ \\ \ \left(  g\rightarrow0\right)\ .
\label{e2100}%
\end{equation}
The amplitude will then depend uniquely on the geodesic distance between the
points $\mathbf{x/}\left\vert \mathbf{x}\right\vert $ and $\mathbf{\bar{x}%
/}\left\vert \mathbf{\bar{x}}\right\vert $ in the transverse hyperbolic space
\textrm{H}$_{3}$, and has the Fourier decomposition (\ref{b100}) given by%
\begin{equation}
\hat{\mathcal{A}}_{\mathrm{planar}}\left(  \mathbf{x},\mathbf{\bar{x}}\right)
\simeq-2\pi i~\int d\nu~\alpha\left(  \nu\right)  ~\Omega_{i\nu}\left(
\mathbf{x},\mathbf{\bar{x}}\right)  ~. \label{e3000}%
\end{equation}
As shown in appendix \ref{app1_20}, the Fourier basis of radial functions
$\Omega_{i\nu}\left(  \mathbf{x},\mathbf{\bar{x}}\right)  $ is conveniently
given by the following integral representation\footnote{In terms of the
geodesic distance $\rho$ in  $\mathrm{H}_{3}$, given by $\cosh\rho=-\left(  \mathbf{x}%
\cdot\mathbf{\bar{x}}\right)  \mathbf{/}\left(  \left\vert \mathbf{x}%
\right\vert \left\vert \mathbf{\bar{x}}\right\vert \right)  $, we have that
$\Omega_{i\nu}=\nu\,\sin\nu\rho/(4\pi^{2}\sinh\rho)$, as discussed in
\cite{Paper4}.}%
\begin{equation}
\Omega_{i\nu}\left(  \mathbf{x},\mathbf{\bar{x}}\right)  =\frac{\nu^{2}}%
{4\pi^{3}}\int_{\partial\mathrm{H}_{3}}d\mathbf{z}_{7}\,\frac{\left(
-\mathbf{x}^{2}\right)  ^{\frac{1+i\nu}{2}}}{\left(  -2\mathbf{x\cdot z}%
_{7}\right)  ^{1+i\nu}}\frac{\left(  -\mathbf{\bar{x}}^{2}\right)
^{\frac{1-i\nu}{2}}}{\left(  -2\mathbf{\bar{x}\cdot z}_{7}\right)  ^{1-i\nu}%
}~, \label{e4000}%
\end{equation}
as shown graphically in figure \ref{fig1}. 

\subsection{Impact factors in position space}

We are now in position to introduce the BFKL formalism in position space, applying it to
the computation of $\hat{\mathcal{A}}_{\mathrm{planar}}\left(  \mathbf{x},\mathbf{\bar{x}}\right)$ 
in the limit $\mathbf{x},\mathbf{\bar{x}}\rightarrow0$ and to leading order in $g^{2}$. The
amplitude $\hat{\mathcal{A}}_{\mathrm{planar}}$ will be given again by
an expression similar to (\ref{e100}), but now the external state impact factors
$V$ and $\bar{V}$ are not labeled by the exchanged transverse momentum $q$, but by
the positions $\mathbf{x}$ and $\mathbf{\bar{x}}$ in the Milne cone $\mathrm{M}$. Therefore we
expect the amplitude $\hat{\mathcal{A}}_{\mathrm{planar}}$ to be given by
an integral of the form
\begin{align}
\hat{\mathcal{A}}_{\mathrm{planar}}\left(  \mathbf{x},\mathbf{\bar{x}}\right)
  \simeq&-\int_{\partial\mathrm{H}_{3}}d\mathbf{z}_{1}d\mathbf{z}_{3}
d\mathbf{z}_{2}d\mathbf{z}_{4}~\label{e600}\\
&  V\left(  \mathbf{x,z}_{1}\mathbf{,z}_{3}\right)  \; F\left(
\mathbf{z}_{1}\mathbf{,z}_{3},\mathbf{z}_{2},\mathbf{z}_{4}\right) \;
\bar{V}\left(  \mathbf{\bar{x},z}_{2}\mathbf{,z}_{4}\right)  ~.\nonumber
\end{align}
Note that we replaced the integrals over the transverse space
$\mathbb{E}^{2}$ with integrals over an arbitrary section $\partial
\mathrm{H}_{3}$ of the light--cone $\partial\mathrm{M}$. 

The form of the
impact factors $V$ and $\bar{V}$ is almost fixed by conformal invariance. In fact,
$V\left(  \mathbf{x,z}_{1}\mathbf{,z}_{3}\right)  $ must scale in
$\mathbf{z}_{1}$ and $\mathbf{z}_{3}$ with weight $2$, in order for the integrals
over $\partial\mathrm{H}_{3}$ to give a conformally invariant result.
Moreover, $V\left(  \mathbf{x,z}_{1}\mathbf{,z}_{3}\right)  $ must be
invariant under rescalings of $\mathbf{x}$ in order to satisfy (\ref{e2100}).
Finally, since $\mathbf{z}_{i}^{2}=0$, there is a unique scale--invariant
conformal cross ratio which can be constructed from $\mathbf{x}$, $\mathbf{z}_{1}$
and $\mathbf{z}_{3}$, given by
\begin{equation}
u=\frac{-\mathbf{x}^{2}~\mathbf{z}_{13}}{\left(  -2\mathbf{x\cdot z}%
_{1}\right)  \left(  -2\mathbf{x\cdot z}_{3}\right)  }~. \label{e602}%
\end{equation}
The impact factor $V$ must then be of the general form%
\[
V\left(  \mathbf{x,z}_{1}\mathbf{,z}_{3}\right)  =\frac{1}{\mathbf{z}_{13}^{2}}\,V(u)  ~.
\]
Let us first note that, since $\mathbf{x}\in\mathrm{M}$ and 
$\mathbf{z}_{1},\mathbf{z}_{3}\in\partial\mathrm{M}$, the cross--ratio $u$ satisfies
\[
0\leq u\leq1~.
\]
This can be shown simply by using $SO\left(  3,1\right)  $ symmetry to rotate
$\mathbf{x}$ and $\mathbf{z}_{3}$ respectively to $\left(  1,1,0\right)  $ and $\left(
1,0,0\right)  $, possibly after an immaterial rescaling. Then, parameterizing
$\mathbf{z}_{1}=\left(  1,z_{1}^{2},z_{1}\right)  $, we obtain $\mathbf{z}%
_{13}=z_{1}^{2}$ and%
\[
u=\frac{z_{1}^{2}}{1+z_{1}^{2}}~.
\]
The infrared finiteness condition (\ref{e200}) may also be written simply as%
\begin{equation}
\int_{0}^{1}\frac{du}{u^{2}}~V(u)  =0~. \label{e2300}%
\end{equation}
In fact, by conformal invariance and scaling, we know that the integral
(\ref{e200}) must be given by%
\[
\int_{\partial\mathrm{H}_{3}}d\mathbf{z}_{1}~V\left(  \mathbf{x,z}%
_{1}\mathbf{,z}_{3}\right)  =c~\frac{-\mathbf{x}^{2}}{\left(  -2\mathbf{x\cdot
z}_{3}\right)  ^{2}}~,
\]
where the constant $c$ can be computed as%
\[
c=\int_{\mathbb{E}^{2}}\frac{dz_{1}}{z_{1}^{4}}\,V\left(  \frac{z_{1}^{2}%
}{1+z_{1}^{2}}\right)  =\pi\int_{0}^{1}\frac{du}{u^{2}}~V(u)  ~.
\]
Similar equations apply to $\bar{V}$.

\subsection{A basis for impact factors}

\begin{figure}[ptb]
\begin{center}
\includegraphics[height=1.8985in]{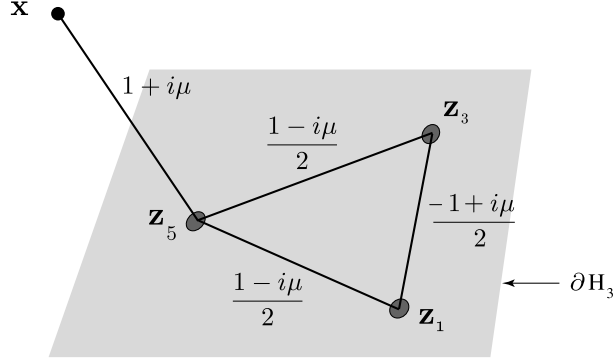}
\end{center}
\caption{Integral representation of the functions $\phi_{\mu}(u)+\phi_{-\mu}( u)$ 
used as a complete basis for the impact factor $V(u)$.}%
\label{fig4}%
\end{figure}

Now we discuss a convenient basis for the functions $V(u)$
satisfying (\ref{e2300}). We shall consider the following
conformal integral%
\begin{equation}
\mu^{2}~c(\mu)  \int_{\partial\mathrm{H}_{3}}d\mathbf{z}_{5}\,
\frac{\left(  -\mathbf{x}^{2}\right)  ^{\frac{1+i\mu}{2}}}{\left(
-2\mathbf{x\cdot z}_{5}\right)  ^{1+i\mu}}~\left(  \frac{\mathbf{z}_{13}%
}{\mathbf{z}_{15}\mathbf{z}_{35}}\right)  ^{\frac{1-i\mu}{2}}, \label{e3200}%
\end{equation}
graphically shown in figure \ref{fig4}, where we defined
\[
c(\mu)  =\frac{1+\mu^{2}}{64\pi^{5}}~\frac{\Gamma^{2}\left(
\frac{1-i\mu}{2}\right)  }{\Gamma\left(  1-i\mu\right)  }~.
\]
By conformal invariance, the integral (\ref{e3200}) is only a
function of the cross--ratio $u$. As shown in appendix \ref{app1_32}, it is
explicitly given by
\begin{equation}
\phi_{\mu}(u)  +\phi_{-\mu}(u)  ~, \label{e5000}%
\end{equation}
where
\[
\phi_{\mu}(u)  =i\pi\mu~c(-\mu) ~u^{\frac{1+i\mu
}{2}}F\left(  \frac{1+i\mu}{2},\frac{1+i\mu}{2},1+i\mu,u\right)  ~,
\]
with $F$ the hypergeometric function ${}_{2}F_{1}$. The functions
$\phi_{\mu}(u)  +\phi_{-\mu}(u)$ are a convenient
basis for the impact factor $V(u)$, which we write as
\begin{equation}
V(u)  =\int d\mu~V(\mu)  ~
\Big[  \phi_{\mu}(u)  +\phi_{-\mu}(u)  \Big]  
=~2\int d\mu~V(\mu)  ~\phi_{\mu}(u)\  , 
\label{e2400}
\end{equation}
where we have chosen
\[
V(\mu)  =V(  -\mu)  ~
\]
without loss of generality. Moreover, we shall use the same label $V$ for the
impact factor both as a function of $u$ and of the transformed variable $\mu$,
with the hope that the difference will be clear from context.

Consider the infrared condition (\ref{e2300}). Since\footnote{The integral diverges at $u=0$ for
real $\mu$ and it is computed by analytically continuing the result obtained for $\mathrm{Im}\mu<-1$.}
\[
\int_{0}^{1}\frac{du}{u^{2}}~\phi_{\mu}(u)  =-\frac{i\mu}{16\pi^{4}}~
\]
is odd in $\mu$, it is clear that $V(u)$ satisfies automatically (\ref{e2300}).

In particular, let us consider $V\left(  u\right)  $ given by a pure power
$u^{\sigma}$. To satisfy the infrared condition (\ref{e2300}), the full
expression must be of the form
\[
V(u)  =u^{\sigma}-\frac{1}{\sigma-1}\,u^{2}\,\delta(u)~.
\]
As shown in appendix \ref{app3}, this corresponds to a transform $V(\mu)$ in
(\ref{e2400}) given by
\[
V(\mu)  =\frac{8\pi^{3}}{1+\mu^{2}}~\frac{\Gamma\left(
\sigma-\frac{1}{2}+\frac{i\mu}{2}\right)  \Gamma\left(  \sigma-\frac{1}%
{2}-\frac{i\mu}{2}\right)  }{\Gamma^{2}\left(  \sigma\right)  }\ .
\]
In particular, we shall see that the relevant impact factor
in section \ref{SecN4} will be
\[
V(u)  =u^{2}\big[ 1 - \delta(u)\big] ~,
\]
corresponding to
\begin{equation}
V(\mu)  =\frac{2\pi^{4}}{\cosh\frac{\pi\mu}{2}}~. \label{r3}%
\end{equation}

\subsection{Computation of the BFKL\ amplitude\label{s200}}

We are now in position to compute the amplitude (\ref{e3000}) starting from
impact factors $V$ and $\bar{V}$. More precisely, we shall show that the new
BFKL integral representation (\ref{e600}) gives an amplitude of the form
(\ref{e3000}) with
\begin{equation}
\alpha(\nu)  =-\frac{i}{4\pi}~V(\nu)  \,
\frac{\tanh\frac{\pi\nu}{2}}{\nu}\,\bar{V}(\nu)  ~. \label{e3100}%
\end{equation}
This will be the main result of this section, showing (\ref{in500}).

\begin{figure}[ptb]
\begin{center}
\includegraphics[height= 1.9034in]{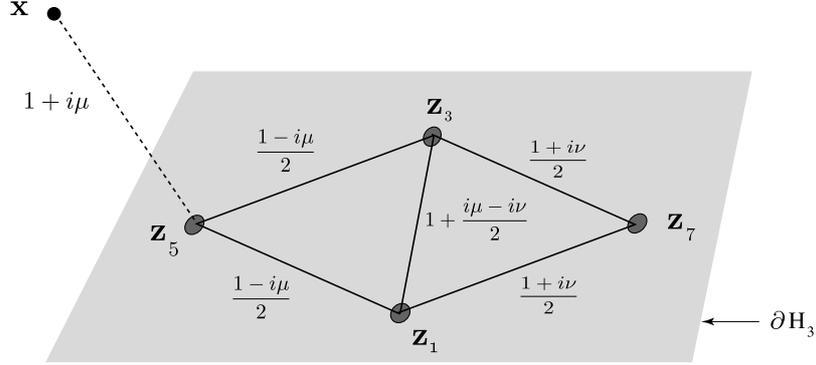}
\end{center}
\caption{Basic overlap between the functions $\phi_{\mu}(u)
+\phi_{-\mu}(u)  $, which are a basis for the impact factor
$V(u)$, and the left part of the $n=0$ component of the BFKL kernel.}%
\label{fig5}
\end{figure}

\begin{figure}[ptb]
\begin{center}
\includegraphics[height= 1.2214in]{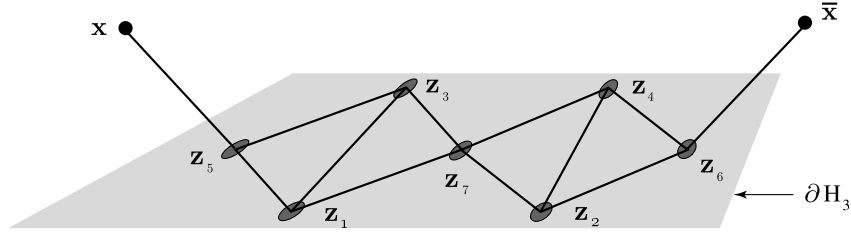}
\end{center}
\caption{Full BFKL amplitude, written as a product of the left and right
impact factors and of the $n=0$ component of the BFKL kernel.}%
\label{fig6}%
\end{figure}

We start by replacing, in the amplitude (\ref{e600}), the $n=0$ part of the
BFKL kernel (\ref{e3300}), thus obtaining
\begin{align}
\hat{\mathcal{A}}_{\mathrm{planar}}\left(  \mathbf{x},\mathbf{\bar{x}}\right)
&  \simeq-\frac{4}{\pi^{2}}\int d\nu~\frac{\nu^{2}}{\left(  1+\nu^{2}\right)
^{2}}\int_{\partial\mathrm{H}_{3}}d\mathbf{z}_{7}\nonumber\\
&  \int_{\partial\mathrm{H}_{3}}d\mathbf{z}_{1}d\mathbf{z}_{3}~V\left(
\mathbf{x,z}_{1}\mathbf{,z}_{3}\right)  ~\left(  \frac{\mathbf{z}_{13}%
}{\mathbf{z}_{17}\mathbf{z}_{37}}\right)  ^{\frac{1+i\nu}{2}}~\label{e3600}\\
&  \int_{\partial\mathrm{H}_{3}}d\mathbf{z}_{2}d\mathbf{z}_{4}~\bar{V}\left(
\mathbf{\bar{x},z}_{2}\mathbf{,z}_{4}\right)  ~\left(  \frac{\mathbf{z}_{24}%
}{\mathbf{z}_{27}\mathbf{z}_{47}}\right)  ^{\frac{1-i\nu}{2}}~.\nonumber
\end{align}
We shall first focus on the second line of this expression. Replacing the
integral representation (\ref{e2400}) for the impact factor $V$, we obtain the
following conformal integral%
\begin{align}
&  \int d\mu~V(\mu)  ~\mu^{2}~c(\mu)  \int_{\partial\mathrm{H}_{3}}
d\mathbf{z}_{5}~\frac{\left(  -\mathbf{x}^{2}\right)^{\frac{1+i\mu}{2}}}
{\left(  -2\mathbf{x\cdot z}_{5}\right)  ^{1+i\mu}}%
~\times\nonumber\\
&  \times\int_{\partial\mathrm{H}_{3}}\frac{d\mathbf{z}_{1}d\mathbf{z}_{3}%
}{\mathbf{z}_{13}^{2}}~\left(  \frac{\mathbf{z}_{13}}{\mathbf{z}%
_{15}\mathbf{z}_{35}}\right)  ^{\frac{1-i\mu}{2}}\left(  \frac{\mathbf{z}%
_{13}}{\mathbf{z}_{17}\mathbf{z}_{37}}\right)  ^{\frac{1+i\nu}{2}%
}~,\label{e3500}%
\end{align}
shown graphically in figure \ref{fig5}. Let us note that the second line of this
expression, highlighted in figure \ref{fig5} with a continuous line, is almost completely
fixed by conformal invariance. It is, in fact, a conformal function 
$f\left(\mathbf{z}_{5},\mathbf{z}_{7}\right)  $ with weights $1-i\mu$ and $1+i\nu$,
respectively in the two entries. Since the only conformal invariant is $\mathbf{z}_{57}$,
the function
$f$ must vanish for $\mu\neq-\nu$ and must be proportional to $\mathbf{z}%
_{57}^{-1-i\nu}$ for $\mu=-\nu$. The second possibility is a contact
$\mathbf{\delta}$--function contribution $\mathbf{\delta}\left(
\mathbf{z}_{5},\mathbf{z}_{7}\right)  $, defined as usual by
\begin{equation}
\int_{\partial\mathrm{H}_{3}}d\mathbf{z}_{7}~\mathbf{\delta}\left(
\mathbf{z}_{5},\mathbf{z}_{7}\right)  ~g\left(  \mathbf{z}_{7}\right)
=g\left(  \mathbf{z}_{5}\right)  ~.\label{e10000}%
\end{equation}
The function $\mathbf{\delta}\left(  \mathbf{z}_{5},\mathbf{z}_{7}\right)  $
is conformally invariant whenever the weights in $\mathbf{z}_{5}$ and
$\mathbf{z}_{7}$ sum to $2$. In fact, if $g$ is of weight $\Delta$, the above integral
is well defined when the weight in $\mathbf{z}_{7}$ is $2-\Delta$ and, 
for (\ref{e10000}) to be satisfied, the weight in
$\mathbf{z}_{5}$ must be $\Delta$. Therefore, the $\mathbf{\delta}$--function
contribution to $f$ can be non--vanishing only for $\mu=\nu$. The exact
integral $f$  has been explicitly computed by Lipatov in \cite{Lipatov},
with the result
\[
\frac{4\pi^{4}}{\nu^{2}}~\mathbf{\delta}\left(  \mathbf{z}_{5},\mathbf{z}%
_{7}\right)  \,\delta(\nu-\mu)  +\frac{4\pi^{3}}{i\nu}%
\frac{c(\nu)  }{c(  -\nu)  }\frac{1}{\mathbf{z}%
_{57}^{1+i\nu}}\,\delta(\nu+\mu)  ~.
\]
We may then complete the computation of (\ref{e3500}), performing the integral
in $\mathbf{z}_{5}$ to obtain
\[
4\pi^{4}\frac{\left(  -\mathbf{x}^{2}\right)  ^{\frac{1+i\nu}{2}}}{\left(
-2\mathbf{x\cdot z}_{7}\right)  ^{1+i\nu}}\int d\mu~V(\mu)\,
c(\mu)  \left[  ~\delta(\nu-\mu)  +\frac{c(\nu)}{c(-\nu)}\,\delta(\nu+\mu)\right]  ~,
\]
where in the second term we have used the conformal integral%
\[
\int_{\partial\mathrm{H}_{3}}d\mathbf{z}_{5}~\frac{\left(  -\mathbf{x}%
^{2}\right)  ^{\frac{1-i\nu}{2}}}{\left(  -2\mathbf{x\cdot z}_{5}\right)
^{1-i\nu}}\frac{1}{\mathbf{z}_{57}^{1+i\nu}}=\frac{i\pi}{\nu}\frac{\left(
-\mathbf{x}^{2}\right)  ^{\frac{1+i\nu}{2}}}{\left(  -2\mathbf{x\cdot z}%
_{7}\right)  ^{1+i\nu}}%
\]
from appendix \ref{app1_21}. Finally, computing the $\mu$ integral and using
the fact that $V(\nu)  =V(-\nu)$ we obtain the
final result for the second line of (\ref{e3600})%
\[
8\pi^{4}\frac{\left(  -\mathbf{x}^{2}\right)  ^{\frac{1+i\nu}{2}}}{\left(
-2\mathbf{x\cdot z}_{7}\right)  ^{1+i\nu}}~V(\nu)\,c(\nu)  ~.
\]
We may carry out an equivalent computation for the second impact factor
$\bar{V}$. Combining the two expressions, we conclude that the
BFKL amplitude (\ref{e3600}), graphically shown in figure \ref{fig6},
is given by%
\begin{align*}
&  -256\pi^{6}\int d\nu~\nu^{2}~V(\nu)  ~\frac{c(\nu)  c(-\nu)  }
{\left(  1+\nu^{2}\right)  ^{2}}~\bar{V}(\nu)  \times\\
&  \times\int_{\partial\mathrm{H}_{3}}d\mathbf{z}_{7}~\frac{\left(
-\mathbf{x}^{2}\right)  ^{\frac{1+i\nu}{2}}}{\left(  -2\mathbf{x\cdot z}
_{7}\right)  ^{1+i\nu}}~\frac{\left(  -\mathbf{\bar{x}}^{2}\right)
^{\frac{1-i\nu}{2}}}{\left(  -2\mathbf{\bar{x}\cdot z}_{7}\right)  ^{1-i\nu}}~.
\end{align*}
Using the fact that
\[
\frac{c(\nu)\,  c(-\nu)  }{\left(  1+\nu^{2}\right)^{2}}=
\frac{1}{4\left(  2\pi\right)  ^{9}}~\frac{\tanh\frac{\pi\nu}{2}}{\nu}~,
\]
together with the integral representation (\ref{e4000}) for the radial Fourier
functions $\Omega_{i\nu}\left(  \mathbf{x,\bar{x}}\right)  $, we obtain the
final result for the amplitude
\[
-\frac{1}{2}\int d\nu~V(\nu)  \,\frac{\tanh\frac{\pi\nu}{2}}{\nu}\,
\bar{V}(\nu) ~~\Omega_{i\nu}\left(  \mathbf{x,\bar{x}%
}\right)  ~,
\]
thus proving (\ref{e3100}).

\subsection{Vanishing of the $n>0$ contributions}

We have previously claimed, without proof, that the unique contribution to the
BFKL\ amplitude (\ref{e600}) comes from the $n=0$ part (\ref{e3300}) of the
complete two--gluon kernel (\ref{e1200}), whenever the external states are
scalar operators. This fact is now almost trivial to show. In fact, the $n>0$
terms would involve, similarly to the discussion in section \ref{s200}, an
overlap integral of the general form (\ref{e3500}). The only difference would
come from the second line of (\ref{e3500}), which would have a $3$--point
coupling at points $\mathbf{z}_{1},\mathbf{z}_{3},\mathbf{z}_{7}$ with a spin
$n\neq0$ state located at $\mathbf{z}_{7}$. The full integral on the
second line of (\ref{e3500}) would then vanish by conservation of transverse
spin, as shown also in \cite{Lipatov}, since it would connect a spin $0$ state
at $\mathbf{z}_{5}$ to a spin $n\neq0$ at $\mathbf{z}_{7}$.

In this paper we  consider only scalar external operators for
simplicity. We could have considered more general external states in various
representations of the $4$--dimensional conformal group. For example, we could
have chosen spin $J$ external states. In this case, the impact factors $V$
would have a non trivial index structure coming from the external operator
$\mathcal{O}_{1}$ at points $\mathbf{x}_{1},\mathbf{x}_{3}$, and the basis
functions (\ref{e3200}) need to be modified to include this extra structure.
It is natural to expect that this will involve contributions of transverse
conformal spin $n\leq2J$ coming from the indices at the two points
$\mathbf{x}_{1},\mathbf{x}_{3}$. This fact was shown in a non--transparent way
in \cite{Munier} for the case $J=1$, which is relevant to interactions with
off--shell photons in deep inelastic scattering processes at small values of
Bjorken $x$.

\section{Impact Factors in $\mathcal{N}=4$ SYM\label{SecN4}}

In this section, we apply the position space BFKL formalism to the computation
of the $\mathcal{N}=4$ SYM $4$--point function%
\[
\left\langle \mathcal{O}_{1}\left(  \mathbf{x}_{1}\right)  \mathcal{O}%
_{1}^{\star}\left(  \mathbf{x}_{3}\right)  \mathcal{O}_{2}\left(
\mathbf{x}_{2}\right)  \mathcal{O}_{2}^{\star}\left(  \mathbf{x}_{4}\right)
\right\rangle \
\]
discussed in section \ref{SecG}. Recall that the operators $\mathcal{O}_{1}$ and 
$\mathcal{O}_{2}$ are given by
\[
\mathcal{O}_{1}=c~\mathrm{Tr}\left(  Z^{2}\right)
\ ,\ \ \ \ \ \ \ \ \ \ \ \ \ \ \ \mathcal{O}_{2}=c\ \mathrm{Tr}\left(
W^{2}\right)  \ ,
\]
with $Z$ and $W$ adjoint complex scalar fields, and are normalized so that
their $2$--point function is
\[
\langle\mathcal{O}_{1}\left(  \mathbf{x}\right)  \mathcal{O}_{1}^{\star
}\left(  \mathbf{y}\right)  \rangle=\langle\mathcal{O}_{2}\left(
\mathbf{x}\right)  \mathcal{O}_{2}^{\star}\left(  \mathbf{y}\right)
\rangle=\frac{1}{\big(  (\mathbf{x}-\mathbf{y})^{2}+i\epsilon\big)^{2}}\ .
\]
In the conventions of appendix \ref{AppN4}, the constant $c$ is given by%
\[
c=\frac{4\pi^{2}}{g_{\mathrm{YM}}^{2}}\frac{\sqrt{2}}{\sqrt{N^{2}-1}}~.
\]
In particular, we shall  compute explicitly the impact factors $V$ and $\bar
{V}$ for the operators $\mathcal{O}_{1}$ and $\mathcal{O}_{2}$ to leading
order in perturbation theory, thus showing (\ref{in501}).

\subsection{Some kinematics}

\begin{figure}[t]
\begin{center}
\includegraphics[
height=1.8474in
]{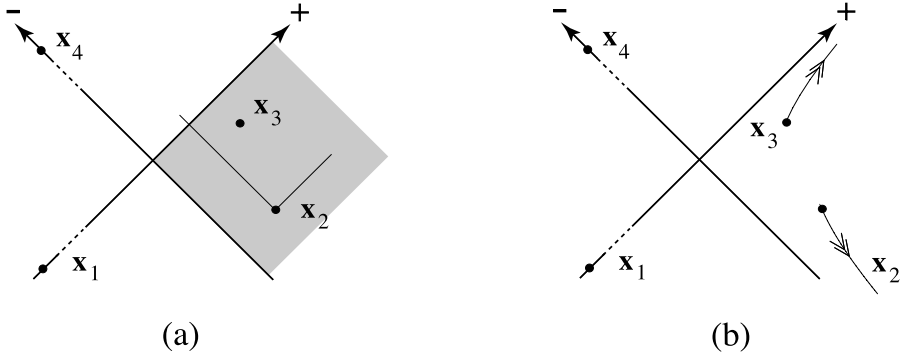}
\end{center}
\caption{Kinematics used for the computation of the impact factors. (a) We choose
$\mathbf{x}_1=(-\infty,0,0)$ and $\mathbf{x}_4=(\infty,0,0)$, and vanishing transverse 
parts of $\mathbf{x}_2,\mathbf{x}_3$. As shown in the text  $\pm x_2^\pm, \pm x_3^\pm >0$, 
with $\mathbf{x}_3$ in the future of $\mathbf{x}_2$. (b) The limit $z,\bar{z}\rightarrow0$, 
with fixed ratio $\bar{z}/z$, described in the text.}
\label{fig9}
\end{figure}

To simplify the computation,  it is
convenient to carefully choose the kinematics.
We shall write $\mathbf{x}=\left(  x^{+},x^{-},x\right)$ to compactly show the light--cone
and transverse components of a vector $\mathbf{x}$.
Following \cite{Paper4}, we choose
\[
\mathbf{x}_{1}=\left(  -s,0,0\right)~,\ \ \ \ \ \ \ \ \ \ \ \ \ \ \ \ \ \ \ \ \ \ \ \mathbf{x}_{4}=\left(
0,s,0\right)  ~,
\]
and we shall consider the limit $s\rightarrow\infty$. The conditions
(\ref{in100}) and (\ref{in100BIS}) then imply that
\[
x_{3}^{+},x_{2}^{+}>0~,\ \ \ \ \ \ \ \ \ \ \ \ \ \ \ x_{3}^{-},x_{2}^{-}<0~,
\]
and that $\mathbf{x}_{3}$ is in the future of $\mathbf{x}_{2}$. In the limit
$s\rightarrow\infty$ the expressions for $\mathbf{x},\mathbf{\bar{x}}$ in
section \ref{SecGeneral} simplify to
\[
\mathbf{x}=\frac{s}{x_{3}^{+}}\left(  s,\frac{1}{s}\,\mathbf{x}_{3}^{2}\,,\,x_{3}\right)~,
\ \ \ \ \ \ \ \ \ \ \ \ \ \ \ \ \ \ 
\mathbf{\bar{x}}=-\frac{1}{sx_{2}^{-}}\left(  s,\frac{1}{s}\,\mathbf{x}_{2}^{2}\,,\,x_{2}\right)~.
\]
Recall that $\mathbf{x},\mathbf{\bar{x}}$ are defined up to the residual
$SO\left(  1,1\right)  \times SO\left(  3,1\right)  $ transverse conformal
symmetry. Therefore, rescaling $\mathbf{x\rightarrow x/}s$ and $\mathbf{\bar{x}%
}\rightarrow s\mathbf{\bar{x}}$, and boosting $x^{\pm}\rightarrow x^{\pm}%
s^{\mp1}$, we obtain the expressions
\[
\mathbf{x}=\frac{1}{x_{3}^{+}}\left(  1,\mathbf{x}_{3}^{2},x_{3}\right)~,
\ \ \ \ \ \ \ \ \ \ \ \ \ \ \ \ \ \ 
\mathbf{\bar{x}}=-\frac{1}{x_{2}^{-}}\left(  1,\mathbf{x}_{2}^{2},x_{2}\right)~,
\]
as in \cite{Paper4}. We can further simplify our computations by choosing
the transverse parts $x_{2},x_{3}$ of the points $\mathbf{x}_{2},\mathbf{x}_{3}$ 
to vanish, so that
\[
\mathbf{x}=\left(  \frac{1}{x_{3}^{+}},-x_{3}^{-},0\right)~,
\ \ \ \ \ \ \ \ \ \ \ \ \ \ \ \ \ \ 
\mathbf{\bar{x}}=\left(  -\frac{1}{x_{2}^{-}},x_{2}^{+},0\right)~.
\]
In this convenient kinematical setup, shown in figure \ref{fig9}a, the
cross--ratios $z,\bar{z}$  read
\[
z=\frac{x_{3}^{-}}{x_{2}^{-}}%
~,~\ \ \ \ \ \ \ \ \ \ \ \ \ \ \ \ \ \ \ \bar{z}=\frac
{x_{2}^{+}}{x_{3}^{+}}~.
\]
The limit $z,\bar{z}\rightarrow0$ with fixed ratio $\bar{z}/z$ can then be
achieved by sending $x_{3}^{+}\rightarrow\infty$, with $x_{3}^{+}x_{3}^{-}$
fixed, and $x_{2}^{-}\rightarrow-\infty$, with $x_{2}^{+}x_{2}^{-}$ fixed, as
shown in figure \ref{fig9}b.

\subsection{Impact factor\label{SecIF}}

\begin{figure}[t]
\begin{center}
\includegraphics[width=12.5cm]{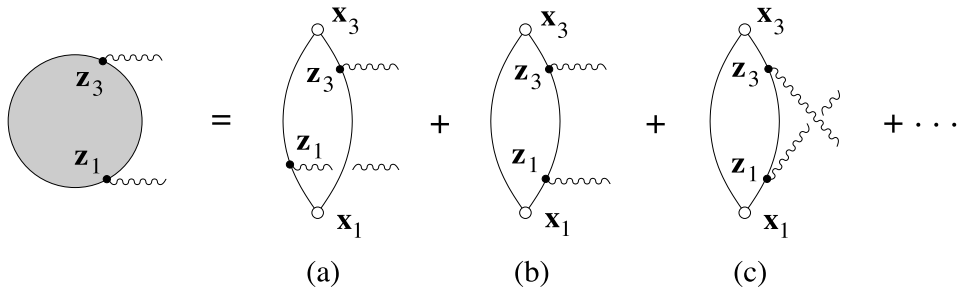}
\end{center}
\caption{Perturbative expansion of the impact factor. Two gluons are emitted
in a color singlet at points $\mathbf{z}_{1}$ and $\mathbf{z}_{3}$, which
become points in transverse space.}%
\label{vertex}%
\end{figure}

\begin{figure}[b]
\begin{center}
\includegraphics[width=12.5cm]{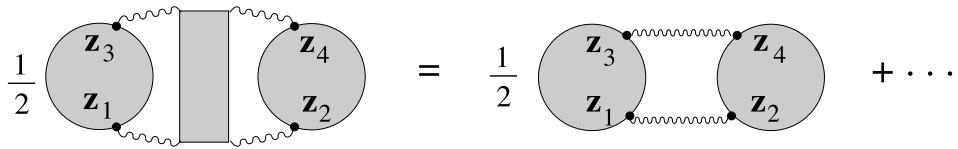}
\end{center}
\caption{Perturbative expansion of the BFKL kernel. The leading term
corresponds to the exchange of a pair of gluons in a color singlet state.}%
\label{pomeron}%
\end{figure}

Let us now compute the impact factor for the external operator $\mathcal{O}%
_{1}$. A similar computation would give the impact factor of $\mathcal{O}_{2}%
$. The leading order diagrams that contribute to the BFKL vertex
$V(\mathbf{x},\mathbf{z}_{1},\mathbf{z}_{3})$ are given in figure
\ref{vertex}, representing the emission of two gluons. The full correlator is
then obtained by connecting both vertices $V$ and $\bar{V}$ with a pomeron
propagator, as described in figure \ref{pomeron}, where the factor of $1/2$ is
the overall symmetry factor of the diagram. To leading order in perturbation theory, the
pomeron propagator is simply given by the exchange of two gluons in a color
singlet state.

First we consider the contribution coming from diagram \ref{vertex}a. Since we
are interested in the reduced amplitude, we must divide the diagram by the two
point function $\langle\mathcal{O}_{1}\left(  \mathbf{x}_{1}\right)
\mathcal{O}_{1}^{\star}\left(  \mathbf{x}_{3}\right)  \rangle$. Fixing for now
the position of the gluons at $\mathbf{z}_{1}$ and $\mathbf{z}_{3}$, the
Feynman rules give
\begin{align*}
&  \left(  -\frac{i}{g_{\mathrm{YM}}^{2}}\right)  ^{2}f_{mna}\,f_{mnb}\left(
\frac{g_{\mathrm{YM}}^{2}}{4\pi^{2}}\right)  ^{4}c^{2}\,\left(
\mathbf{x}_{1}-\mathbf{x}_{3}\right)  ^{4}\\
&  \,\left(  \frac{1}{\left(  \mathbf{x}_{1}-\mathbf{z}_{1}\right)
^{2}+i\epsilon}\,{\overleftrightarrow{\partial}}_{\hspace{-0.15cm}z_{1}^{\mu}%
}\,\frac{1}{\left(  \mathbf{x}_{3}-\mathbf{z}_{1}\right)  ^{2}+i\epsilon
}\right)  \ \ \\
&  \left(  \frac{1}{\left(  \mathbf{x}_{1}-\mathbf{z}_{3}\right)
^{2}+i\epsilon}\,{\overleftrightarrow{\partial}}_{\hspace{-0.15cm}z_{3}^{\nu}%
}\,\frac{1}{\left(  \mathbf{x}_{3}-\mathbf{z}_{3}\right)  ^{2}+i\epsilon
}\right)  \ ,
\end{align*}
where $\mu,\nu$ and $a,b$ are the spacetime and color indices of the gluons emitted at
$\mathbf{z}_{1},\mathbf{z}_{3}$. We remark that for now $\mathbf{z}_{1}$ and 
$\mathbf{z}_{3}$ are points in the physical $4$--dimensional Minkowski spacetime. Later
on in the computation these points will collapse to transverse space 
$\mathbb{E}^{2}$, and we shall used the embedding formalism described in section
\ref{SecConfInv}.
Simplifying the overall constant in the above expression, we
obtain
\begin{align}
&  -\frac{2N\delta_{ab}}{(2\pi)^{4}\left(  N^{2}-1\right)  }\,\left(
\frac{\left(  \mathbf{x}_{1}-\mathbf{x}_{3}\right)  ^{2}}{\left(
\mathbf{x}_{1}-\mathbf{z}_{1}\right)  ^{2}+i\epsilon}\,{\overleftrightarrow
{\partial}}_{\hspace{-0.15cm}z_{1}^{\mu}}\,\frac{1}{\left(  \mathbf{x}%
_{3}-\mathbf{z}_{1}\right)  ^{2}+i\epsilon}\right)  \ \ \nonumber\\
&  \left(  \frac{\left(  \mathbf{x}_{1}-\mathbf{x}_{3}\right)  ^{2}}{\left(
\mathbf{x}_{1}-\mathbf{z}_{3}\right)  ^{2}+i\epsilon}\,{\overleftrightarrow
{\partial}}_{\hspace{-0.15cm}z_{3}^{\nu}}\,\frac{1}{\left(  \mathbf{x}%
_{3}-\mathbf{z}_{3}\right)  ^{2}+i\epsilon}\right)  \ , \label{eq:V1}%
\end{align}
which represents the emission at $\mathbf{z}_{1}$ and $\mathbf{z}_{3}$ of two
gluons in a color singlet, respectively with polarizations $\mu$ and $\nu$.

As claimed in the previous section, the perturbative computation simplifies
considerably if we choose the external kinematics using conformal invariance
to set $\mathbf{x}_{1}\rightarrow\left(  -\infty,0,0\right)  $ and
$\mathbf{x}_{3}\rightarrow\left(  x_{3}^{+},x_{3}^{-},0\right)  $. Then, the
term in brackets in the first line of (\ref{eq:V1}) becomes
\[
\frac{x_{3}^{-}}{z_{1}^{-}-i\epsilon}\,\overleftrightarrow{\partial}%
_{\hspace{-0.15cm}z_{1}^{\mu}}\,\frac{1}{-\left(  x_{3}^{+}-z_{1}^{+}\right)
\left(  x_{3}^{-}-z_{1}^{-}\right)  +z_{1}^{2}+i\epsilon}\ .
\]
A similar expression can be obtained for the other bracket with $\mathbf{z}%
_{1}$ replaced by $\mathbf{z}_{3}$. Since the BFKL kinematical limit
corresponds to $x_{3}^{+}$ large with the product $x_{3}^{+}x_{3}^{-}$ held
fixed, this last expression is dominated by the derivative with $\mu=-$, with
the leading result
\begin{equation}
\frac{x_{3}^{-}}{z_{1}^{-}-i\epsilon}\,\overleftrightarrow{\partial}%
_{\hspace{-0.15cm}z_{1}^{-}}\,\frac{1}{x_{3}^{+}\left(  z_{1}^{-}-x_{3}%
^{-}\right)  +z_{1}^{2}+i\epsilon}\ . \label{eq:impact}%
\end{equation}
As expected, the emitted gluons have polarization $\mu=\nu=-$. The computation
of the impact factor for the external operator $\mathcal{O}_{2}$ on the other
side of the graph is analogous to that of $\mathcal{O}_{1}$, representing the
emission of gluons at $\mathbf{z}_{2}$ and $\mathbf{z}_{4}$. In this case we
set $\mathbf{x}_{4}\rightarrow\left(  0,+\infty,0\right)  $ and $\mathbf{x}%
_{2}\rightarrow\left(  x_{2}^{+},x_{2}^{-},0\right)  $, and then take the BFKL
kinematical limit of large negative $x_{2}^{-}$ with $x_{2}^{-}x_{2}^{+}$
fixed. The emitted gluons will have polarization $\bar{\mu}=\bar{\nu}=+$.

To identify the impact factors and the BFKL kernel one needs to integrate over
the internal vertices $\mathbf{z}_{1}$, $\mathbf{z}_{2}$, $\mathbf{z}_{3}$ and
$\mathbf{z}_{4}$, and to add the gluon propagators, as described by figure
\ref{pomeron}. Considering, for example, the vertex at $\mathbf{z}_{1}$, we
shall split the integration in transverse and 
light--cone directions according to
\[
\int d\mathbf{z}_{1}=\int dz_{1}\,dz_{1}^{-}\,\frac{dz_{1}^{+}}{2}\ .
\]
In the BFKL kinematical limit, the external scalar lines are almost on-shell,
while the exchanged gluons are off-shell. When computing the full diagram and
integrating over $z_{1}^{-}$, the residues at the poles in expression
(\ref{eq:impact}) are dominant with respect to the residues at the poles in
the gluon propagators, as we take $x_{3}^{+}$ large with fixed product
$x_{3}^{+}x_{3}^{-}$. Putting together equations (\ref{eq:V1}) and
(\ref{eq:impact}) and dropping the color factor $\delta_{ab}$, we conclude
that the contribution of the diagram in figure \ref{vertex}a to the impact
factor is given by
\begin{align*}
&  -\frac{2N}{\left(  2\pi\right)  ^{4}\left(  N^{2}-1\right)  }\,\left(
x_{3}^{-}\right)  ^{2}\int dz_{1}^{-}\frac{1}{z_{1}^{-}-i\epsilon
}\,\overleftrightarrow{\partial}_{\hspace{-0.15cm}z_{1}^{-}}\,\frac{1}%
{x_{3}^{+}\left(  z_{1}^{-}-x_{3}^{-}\right)  +z_{1}^{2}+i\epsilon}\\
&  \int dz_{3}^{-}\frac{1}{z_{3}^{-}-i\epsilon}\,\overleftrightarrow{\partial
}_{\hspace{-0.15cm}z_{3}^{-}}\,\frac{1}{x_{3}^{+}\left(  z_{3}^{-}-x_{3}%
^{-}\right)  +z_{3}^{2}+i\epsilon}\ ,
\end{align*}
corresponding to the emission of two gluons in a color singlet, located at
$z_{1}$ and $z_{2}$ in transverse space and with polarization $\mu=\nu=-$.
These integrals are easily computed by deforming the contour of integration,
with the result
\begin{equation}
\frac{2N}{\pi^{2}\left(  N^{2}-1\right)  }\frac{-x_{3}^{+}x_{3}^{-}}{\left(
-x_{3}^{+}x_{3}^{-}+z_{1}^{2}\right)  ^{2}}\frac{-x_{3}^{+}x_{3}^{-}}{\left(
-x_{3}^{+}x_{3}^{-}+z_{3}^{2}\right)  ^{2}}\ . \label{e603}%
\end{equation}
Note that, after integrating in $z_{1}^{-}$ and $z_{3}^{-}$, and taking the
BFKL limit $x_{3}^{+}\rightarrow\infty$, the resulting expression is
independent of the other light--cone variables $z_{1}^{+}$ and $z_{3}^{+}$. The
expression depends only on the gluon positions $z_{1},z_{3}$ in transverse
space $\mathbb{E}^{2}$. Recalling from section \ref{SecConfInv} that explicit
transverse conformal invariance is rendered manifest by considering the usual
transverse space $\mathbb{E}^{2}$ as the canonical Poincar\'{e} slice of the
light--cone $\partial\mathrm{M}$, we set $\mathbf{z}_{i}=\left(  1,z_{i}^{2},z_{i}\right)$. 
Note that we use the same label $\mathbf{z}_{i}$ both
for the original position of the gluons and for the points of the Poincar\'{e}
slice. This slight abuse of notation is justified by the fact that the
relevant transverse parts coincide. It is then immediate to show that the 
crossratio $u$ in (\ref{e602}) is given
by
\begin{equation}
u=\frac{-x_{3}^{+}x_{3}^{-}~\left(  z_{1}-z_{3}\right)^{2}}
{\left(-x_{3}^{+}x_{3}^{-}+z_{1}^{2}\right)  
\left(  -x_{3}^{+}x_{3}^{-}+z_{3}^{2}\right)}~, 
\label{e604}
\end{equation}
so that expression (\ref{e603}) can be finally written as
\begin{equation}
\frac{1}{\mathbf{z}_{13}^{\,2}}\,\frac{2N}{\pi^{2}\left(  N^{2}-1\right)
}\,u^{2}~, \label{r1}
\end{equation}
where we recall that $\mathbf{z}_{ij}=-2\mathbf{z}_{i}\cdot\mathbf{z}%
_{j}=\left(  z_{i}-z_{j}\right)  ^{2}$.

Before we compute the contribution to the impact factor of the remaining
diagrams in figure \ref{vertex}, let us consider the BFKL kernel. In the above
computation we saw that the residues of the poles at $z_{1}^{-}=z_{3}^{-}=0$
and at $z_{2}^{+}=z_{4}^{+}=0$ are independent of the other light-cone
integration variables $z_{1}^{+}$, $z_{3}^{+}$, $z_{2}^{-}$ and $z_{4}^{-}$.
Therefore, when computing the full diagram, we can move these integrals to the
gluon propagators. It is then clear that the leading order BFKL propagator, as
represented in figure \ref{pomeron}, is given by
\begin{equation}
\frac{1}{2}\,\int\frac{dz_{1}^{+}}{2}\frac{dz_{2}^{-}}{2}\,D_{a\bar{a}}%
^{-+}(\mathbf{z}_{1},\mathbf{z}_{2})\ \int\frac{dz_{3}^{+}}{2}\frac{dz_{4}%
^{-}}{2}\,D_{a\bar{a}}^{-+}(\mathbf{z}_{3},\mathbf{z}_{4})\ , \label{twogluon}%
\end{equation}
where the spacetime gluon propagators 
$D_{a\bar{a}}^{\mu\nu}(\mathbf{z}_i,\mathbf{z}_j)$ are
computed at the above poles $z_{1}^{-}=z_{3}^{-}=0$ and $z_{2}^{+}=z_{4}%
^{+}=0$. The overall factor of $1/2$ comes from the symmetry factor of the
diagram, while the factors of $1/2$ inside the integration come from the
measure. A simple computation, using%
\[
\int\frac{dz^{+}dz^{-}}{2}\frac{1}{\left(  -z^{+}z^{-}+z^{2}+i\epsilon\right)
}=-i\pi\ln z^{2}~,
\]
gives the transverse gluon propagators\footnote{The result is independent of
the gauge choice, since the gluon propagators have zero longitudinal momenta
and have $-+$ polarization.}
\[
-\frac{g_{\mathrm{YM}}^{4}}{\left(  8\pi\right)  ^{2}}\left(  N^{2}-1\right)
\ 2\ln\left(  {z}_{1}-z_{2}\right)  ^{2}\ln\left(  {z}_{3}-z_{4}\right)
^{2}\ .
\]
The full amplitude has now the BFKL structure (\ref{e600}). The minus sign of
(\ref{e600}) corresponds to the sign of the previous equation. Moreover, 
to match the convention (\ref{e300}) for the two--gluon leading
propagator, we shall multiply, at the end of the computation, the graphs in
figure \ref{vertex} used to compute the impact factor by%
\begin{equation}
\frac{g_{\mathrm{YM}}^{2}}{8\pi}N\sqrt{N^{2}-1}~, \label{e1234}%
\end{equation}
where the extra factor of $N$ comes from our convention on planar amplitudes
(\ref{eq1111}) which explicitly shows an overall factor of $N^{-2}$.

Now we compute the contribution to the impact factor of the diagram in figure
\ref{vertex}b
\begin{align*}
&  \left(  -\frac{i}{g_{\mathrm{YM}}^{2}}\right)  ^{2}f_{mna}\,f_{nmb}\left(
\frac{g_{\mathrm{YM}}^{2}}{4\pi^{2}}\right)  ^{4}c^{2}\,\left(
\mathbf{x}_{1}-\mathbf{x}_{3}\right)  ^{2}\\
&  \,\frac{1}{\left(  \mathbf{x}_{1}-\mathbf{z}_{1}\right)  ^{2}+i\epsilon
}\,{\overleftrightarrow{\partial}}_{\hspace{-0.15cm}z_{1}^{\mu}}\,\frac
{1}{\left(  \mathbf{z}_{1}-\mathbf{z}_{3}\right)  ^{2}+i\epsilon
}\,{\overleftrightarrow{\partial}}_{\hspace{-0.15cm}z_{3}^{\nu}}\,\frac
{1}{\left(  \mathbf{x}_{3}-\mathbf{z}_{3}\right)  ^{2}+i\epsilon}\ ,
\end{align*}
which in the BFKL kinematical limit simplifies to
\[
\frac{2N\delta_{ab}}{(2\pi)^{4}\left(  N^{2}-1\right)  }\,\frac{x_{3}^{-}%
}{x_{3}^{+}}\,\,\frac{1}{z_{1}^{-}-i\epsilon}\,{\overleftrightarrow{\partial}%
}_{\hspace{-0.15cm}z_{1}^{\mu}}\,\frac{1}{\left(  \mathbf{z}_{1}%
-\mathbf{z}_{3}\right)  ^{2}+i\epsilon}\,{\overleftrightarrow{\partial}%
}_{\hspace{-0.15cm}z_{3}^{\nu}}\,\frac{1}{z_{3}^{-}-x_{3}^{-}+\frac{z_{3}^{2}%
}{x_{3}^{+}}+i\epsilon}\ .
\]
If we write, in the full diagram, the gluon propagators in the Landau gauge,
after integrating by parts we may act with the derivatives only on the
internal scalar line, with the result
\begin{align}
&  -\frac{8N\delta_{ab}}{(2\pi)^{4}\left(  N^{2}-1\right)  }\,\frac{x_{3}^{-}
}{x_{3}^{+}}\,\frac{1}{z_{1}^{-}-i\epsilon}\,\frac{1}{z_{3}^{-}-x_{3}
^{-}+\frac{z_{3}^{2}}{x_{3}^{+}}+i\epsilon}\ \label{eq:V2}\\
&  \,\left(  {\partial}_{z_{1}^{\mu}}\,{\partial}_{z_{3}^{\nu}}\,\frac
{1}{\left(  \mathbf{z}_{1}-\mathbf{z}_{3}\right)  ^{2}+i\epsilon}\right)\ .
\nonumber
\end{align}
The impact factor is then computed after integrating this expression in
$z_{1}^{-}$ and $z_{3}^{-}$. The corresponding residues dominate the residues
at the poles of the gluon propagators. First we note that there are singularities in the
previous equation for
\[
\left\{
\begin{array}
[c]{l}%
z_{1}^{-}=i\epsilon\\
z_{1}^{-}=z_{3}^{-}-\frac{\left(  z_{1}-z_{3}\right)  ^{2}}{z_{3}^{+}%
-z_{1}^{+}}\mp i\epsilon
\end{array}
\right.  \ ,\ \ \ \ \ \ \ \ \ \ \left\{
\begin{array}
[c]{l}%
z_{3}^{-}=x_{3}^{-}-\frac{z_{3}^{2}}{x_{3}^{+}}-i\epsilon\\
z_{3}^{-}=z_{1}^{-}+\frac{\left(  z_{1}-z_{3}\right)  ^{2}}{z_{3}^{+}%
-z_{1}^{+}}\pm i\epsilon
\end{array}
\right.  \ ,
\]
where the upper and lower signs correspond, respectively, to $z_{3}^{+}%
>z_{1}^{+}$ and $z_{3}^{+}<z_{1}^{+}$. It is then clear that the $z_{1}^{-}$
and $z_{3}^{-}$ integrals are non-vanishing only for $z_{3}^{+}>z_{1}^{+}$,
which has the physical interpretation of ordering the interaction vertices in
light-cone time. Therefore, we may deform the $z_{1}^{-}$ and $z_{3}^{-}$
integrals in the upper and lower half plane, respectively, picking the
contributions of the poles at $z_{1}^{-}=0$ and at $z_{3}^{-}=x_{3}^{-}%
-z_{3}^{2}/x_{3}^{+}$. To compute the relevant residues, let us first note
that, in the BFKL kinematical regime, the pole in $z_{3}^{-}$ satisfies
$z_{3}^{-}\rightarrow0$ and one needs to keep only the dominant term in this
limit. A simple computation shows that again gluons with polarizations
$\mu=-$ and $\nu=-$ give the dominant term. In particular, at the poles we have
\begin{align*}
{\partial}_{z_{1}^{-}}\,{\partial}_{z_{3}^{-}}\,\frac{1}{\left(
\mathbf{z}_{1}-\mathbf{z}_{3}\right)  ^{2}}  &  =-\frac{2}{\left(  z_{3}%
^{-}\right)  ^{2}}\,\frac{\left(  z_{3}^{-}\left(  z_{3}^{+}-z_{1}^{+}\right)
\right)  ^{2}}{\left(  -z_{3}^{-}\left(  z_{3}^{+}-z_{1}^{+}\right)  +\left(
z_{1}-z_{3}\right)  ^{2}\right)  ^{3}}\\
&  \rightarrow-\frac{\pi}{\left(  z_{3}^{-}\right)  ^{2}}\,\,\delta
^{(2)}\left(  z_{1}-z_{3}\right)  \ ,
\end{align*}
where the last limit is obtained for $z_{3}^{-}=x_{3}^{-}-z_{3}^{2}/x_{3}%
^{+}\rightarrow0$\ using a standard representation of the $\delta
$--function\footnote{For $\mu,\nu$ given by $-,i$ and $i,j$, there are terms
in $\partial_{z_{1}^{\mu}}\partial_{z_{3}^{\nu}}(\mathbf{z}_{1}-\mathbf{z}%
_{3})^{-2}$ which are also of order $(z_{3}^{-})^{-2}$. Such terms are
proportional to $(z_{1}^{i}-z_{3}^{i})\,\delta^{(2)}(z_{1}-z_{3})$ and
therefore vanish.}. We may now return to the computation of the impact factor
in (\ref{eq:V2}), integrating over $z_{1}^{-}$ and $z_{3}^{-}$ we obtain
(dropping again the color factor $\delta_{ab}$ already included in
the two--gluon kernel (\ref{twogluon}))
\begin{equation}
-\frac{2N}{\pi\left(  N^{2}-1\right)  }\,\frac{-x_{3}^{+}x_{3}^{-}}{\left(
-x_{3}^{+}x_{3}^{-}+z_{3}^{2}\right)  ^{2}}\,~\delta^{(2)}\left(  z_{1}%
-z_{3}\right)  \ . \label{e4321}%
\end{equation}
Again this result does not depend of $z_{1}^{+}$ and $z_{3}^{+}$ so that, when
computing the full diagram, their integration can be moved to the gluon
propagators in (\ref{twogluon}). Here one needs to be careful because the
contribution of this diagram gives the restriction $z_{3}^{+}>z_{1}^{+}$ to
the gluon integration. However, repeating the same arguments for the diagram
in figure \ref{vertex}c, we recover the whole integration domain. The
contribution to the impact factor of the diagrams in figure \ref{vertex}b and
\ref{vertex}c is then given by (\ref{e4321}).

Defining the delta function along a radial coordinate in $\mathbb{E}^{2}$ as
\[
\delta^{(2)}(z)=\frac{1}{\pi}\,\delta(r^{2})\ ,\ \ \ \ \ \ \ \ \int
_{0}^{\infty}d(r^{2})\,\delta\left(  r^{2}\right)  =1\ ,
\]
and using the explicit expression for $u$ in (\ref{e604}), we have that%
\[
\delta\left(  u\right)  =\pi~\frac{\left(  -x_{3}^{+}x_{3}^{-}+z_{3}%
^{2}\right)  ^{2}}{-x_{3}^{+}x_{3}^{-}}~\,\delta^{(2)}\left(  z_{1}%
-z_{3}\right)  ~,
\]
so that (\ref{e4321}) reads
\begin{equation}
-\frac{1}{\mathbf{z}_{13}^{\,2}}\frac{2N}{\pi^{2}\left(  N^{2}-1\right)}\,
u^{2}\,\delta\left(  u\right)  ~. \label{r2}%
\end{equation}

Finally, we add the contributions (\ref{r1}) and (\ref{r2}) from all diagrams in
figure \ref{vertex} and multiply by (\ref{e1234}) to obtain the correctly
normalized impact factor. Taking the large $N$ limit we obtain
\[
V(u)=\frac{g^{2}}{4\pi^{3}}\,u^{2}\,\big[  1-\delta(u)\big]  \ ,
\]
where we recall that $g^{2}=g_{\mathrm{YM}}^{2}N$ is the 't Hooft coupling.
Note that the above expression satisfies the infrared finiteness condition
(\ref{e2300}). Using (\ref{r3}), this corresponds to%
\[
V(\mu)  =\frac{\pi~g^{2}}{2}\frac{1}{\cosh\frac{\pi\mu}{2}}~,
\]
thus confirming equation (\ref{in501}).

Let us conclude by quoting a simple extension of the result above which we
prove in appendix \ref{AppImp}. We could have considered the more general
operator
\[
\mathcal{O}_{1}=c_{L}~\mathrm{Tr}\left(  Z^{L}\right)  ~,
\]
where again $c_{L}$ is chosen so that the $2$--point function $\left\langle
\mathcal{O}_{1}\left(  \mathbf{x}\right)  \mathcal{O}_{1}^{\star}\left(
\mathbf{y}\right)  \right\rangle $ is normalized to $\left\vert \mathbf{x}%
-\mathbf{y}\right\vert ^{-2L}$. One may compute the corresponding leading
order impact factor $V(u)$ quite easily. In fact, the spacetime
part of the computation is independent of $L$ and the unique difference is
related to the color factors. A careful analysis shows that the impact factor
in this case is given by%
\begin{equation}
V(u)=\frac{g^{2}L}{8\pi^{3}}\,u^{2}\,\big[  1-\delta(u)\big]  ~.\label{imL}%
\end{equation}


\section*{Acknowledgments}

LC is funded by the \textit{Museo Storico della Fisica e Centro Studi e 
Ricerche "Enrico Fermi"}. LC\ is partially funded by INFN, by the MIUR--PRIN contract 2005--024045--002, by the EU contracts MRTN--CT--2004--005104. JP is funded by the FCT fellowship 
SFRH/BPD/34052/2006. JP and MC are partially funded by the FCT--CERN grant 
POCI/FP/63904/2005. \emph{Centro de F\'{\i}sica do Porto} is partially funded by FCT through the POCI program. This research was supported in part by the National Science Foundation under Grant No. NSF PHY05-51164.


\vfill

\eject

\appendix

\section{Conformal Integrals\label{app1}}

\subsection{General integrals}

We shall work with vectors $\mathbf{x}=\left(  x^{+},x^{-},x\right)  $ in
$\left(  d+2\right)  $--dimensional Minkowski space $\mathbb{M}^{d+2}$ with
norm $\mathbf{x}^{2}=-x^{+}x^{-}+x\cdot x$, and we define, as in the main
text, the usual subspaces%
\begin{align*}
&  \mathrm{M}\text{ future Milne wedge ,}\\
&  \mathrm{H}_{d+1}\subset\mathrm{M}\text{ hyperbolic space of points with
}\mathbf{w}^{2}=-1~,\\
&  \partial\mathrm{M}\text{ future light--cone of points with }\mathbf{w}%
^{2}=0~,\\
&  \partial\mathrm{H}_{d+1}\text{ coiche of arbitrary slice of the light-rays
in }\partial\mathrm{M~.}%
\end{align*}
Let us consider first the following conformal integrals\footnote{The
normalization chosen for the $D$--functions differs by a factor $2/\Gamma
\left(  \frac{\Delta-d}{2}\right)  $ from the one chosen in \cite{Paper1}.}%
\[
D\left(  \mathbf{w}_{1},\cdots,\mathbf{w}_{n}\right)  =\frac{2}{\Gamma\left(
\frac{\Delta-d}{2}\right)  }~\int_{\mathrm{H}_{d+1}}\mathbf{~\widetilde
{d\mathbf{y}}~\ }%
{\textstyle\prod\nolimits_{i}}
~\frac{1}{\left(  -2\mathbf{y\cdot w}_{i}\right)  ^{\Delta_{i}}}~,
\]
where the points $\mathbf{w}_{i}$ are generically in $\mathrm{M}$, or on the
boundary $\partial\mathrm{M}$, and carry weight $\Delta_{i}$, and where we have
defined
\[
\Delta=%
{\textstyle\sum\nolimits_{i}}
\Delta_{i}~.
\]
The above integral converges for
\begin{align}
\operatorname{Re}\Delta &  >d~,\nonumber\\
\operatorname{Re}\Delta_{i}  &  <\operatorname{Re}%
{\textstyle\sum\nolimits_{j\neq i}}
\Delta_{i}~\ \ \ \ \ \ \text{if }\mathbf{w}_{i}\in\partial\mathrm{M~,}
\label{a800}%
\end{align}
and admits the following Feynman parameter representation \cite{Paper1}%
\begin{equation}
D\left(  \mathbf{w}_{i}\right)  =\frac{2\pi^{\frac{d}{2}}}{%
{\textstyle\prod\nolimits_{i}}
\Gamma\left(  \Delta_{i}\right)  }\int\mathbf{~}%
{\textstyle\prod\nolimits_{i}}
dt_{i}~\ t_{i}^{\Delta_{i}-1}~e^{\mathbf{-}\frac{1}{2}\sum_{i,j}t_{i}%
t_{j}~\mathbf{w}_{ij}}~, \label{a200}%
\end{equation}
with $\mathbf{w}_{ij}=-2\mathbf{w}_{i}\cdot\mathbf{w}_{j}$. We will be more
interested in the closely related conformal integral%
\begin{equation}
\tilde{D}\left(  \mathbf{w}_{1},\cdots,\mathbf{w}_{n}\right)  =\int
_{\partial\mathrm{H}_{d+1}}\mathbf{~\widetilde{d\mathbf{z}}~\ }%
{\textstyle\prod\nolimits_{i}}
~\frac{1}{\left(  -2\mathbf{z\cdot w}_{i}\right)  ^{\Delta_{i}}}~,
\label{a100}%
\end{equation}
where we demand that%
\[
\Delta=d~
\]
in order for the result to be conformally invariant. Whenever a point
$\mathbf{w}_{i}$ is on the boundary $\partial\mathrm{M}$, the convergence of
the integral (\ref{a100}) is again ensured by (\ref{a800}), which can also be
written as $\operatorname{Re}\Delta_{i}<d/2$. To compute the integral
(\ref{a100}), we choose the Poincar\'{e} slice for $\partial\mathrm{H}_{d+1}$,
given by $\mathbf{z}=\left(  z^{2},1,z\right)  $, so that
\[
\int_{\partial\mathrm{H}_{d+1}}\widetilde{d\mathbf{z}}\rightarrow
\int_{\mathbb{E}^{d}}dz~.
\]
Using the usual Schwinger representation for the propagators $\left(
-2\mathbf{z\cdot w}_{i}\right)  ^{-\Delta_{i}}$ we obtain%
\[
\frac{1}{%
{\textstyle\prod\nolimits_{i}}
\Gamma\left(  \Delta_{i}\right)  }\int\mathbf{~}%
{\textstyle\prod\nolimits_{i}}
dt_{i}~\ t_{i}^{\Delta_{i}-1}~\int_{\mathbb{E}^{d}}dz~e^{2\mathbf{W\cdot z}%
}~,
\]
where we defined $\mathbf{W=}%
{\textstyle\sum\nolimits_{i}}
\mathbf{w}_{i}t_{i}$. Since $2\mathbf{W\cdot z=}-W^{+}-W^{-}%
z^{2}+2W\cdot z$, the integral over $\mathbb{E}^{d}$ in $z$ is gaussian and
may be evaluated, with the result%
\[
\frac{\pi^{\frac{d}{2}}}{%
{\textstyle\prod\nolimits_{i}}
\Gamma\left(  \Delta_{i}\right)  }\int\mathbf{~}%
{\textstyle\prod\nolimits_{i}}
dt_{i}~\ t_{i}^{\Delta_{i}-1}\left(  W^{-}\right)  ^{-\frac{d}{2}}%
e^{\frac{\mathbf{W}^{2}}{W^{-}}}~.
\]
Finally, changing variables $t_{i}\rightarrow t_{i}W^{-}$,
with $%
{\textstyle\prod\nolimits_{i}}
dt_{i}\rightarrow2%
{\textstyle\prod\nolimits_{i}}
dt_{i}\left(  W^{-}\right)  ^{2}$, we obtain exactly the same expression
(\ref{a200}) for the functions $D\left(  \mathbf{w}_{i}\right)  $, with the
restriction $\Delta=d$. From now on we shall therefore drop the\ tilde. 

Let us conclude by recalling that, if we take $m$ of the $n$ points $\mathbf{w}_{i}$ to
live on future light--cone $\partial\mathrm{M}$, the function $D$ depends in
general on
\[
\frac{1}{2}\,n\left(  n-1\right)  -m
\]
independent cross--ratios.

\subsection{Two point function $n=2$, $m=1$\label{app1_21}}

This is the simplest case, with no cross-ratios. Assuming that $\mathbf{w}%
_{2}\in\partial\mathrm{M}$ and $\operatorname{Re}\Delta_{2}<\operatorname{Re}%
\Delta_{1}$ we have that%
\[
D\left(  \mathbf{w}_{1},\mathbf{w}_{2}\right)  =\pi^{\frac{d}{2}}\frac
{\Gamma\left(  \frac{\Delta_{1}-\Delta_{2}}{2}\right)  }{\Gamma\left(
\Delta_{1}\right)  }\cdot~\frac{\left\vert \mathbf{w}_{1}\right\vert
^{\Delta_{2}-\Delta_{1}}}{\mathbf{w}_{12}^{\Delta_{2}}}~,
\]
where the overall normalization is computed from the integral
\[
\frac{2\pi^{\frac{d}{2}}}{\Gamma\left(  \Delta_{1}\right)  \Gamma\left(
\Delta_{2}\right)  }\int\mathbf{~}dt_{1}dt_{2}~\ t_{1}^{\Delta_{1}-1}%
t_{2}^{\Delta_{2}-1}~e^{\mathbf{-}t_{1}t_{2}-t_{1}^{2}}~.
\]

\subsection{Two point function $n=2$, $m=0$\label{app1_20}}

In this case we have one independent cross--ratio, which we choose to be given
by%
\[
u=\frac{1}{2}-\frac{1}{4}\frac{\mathbf{w}_{12}}{\left\vert \mathbf{w}%
_{1}\right\vert \left\vert \mathbf{w}_{2}\right\vert }~.
\]
We then have that
\[
D\left(  \mathbf{w}_{1},\mathbf{w}_{2}\right)  =\frac{1}{\left\vert
\mathbf{w}_{1}\right\vert ^{\Delta_{1}}\left\vert \mathbf{w}_{2}\right\vert
^{\Delta_{2}}}D_{2}(u)~,
\]
where
\begin{equation}
D_{2}(u)  = \frac{2\pi^{\frac{d}{2}}}{\Gamma\left(  \Delta
_{1}\right)  \Gamma\left(  \Delta_{2}\right)  }\int\mathbf{~}\frac
{dt_{1}dt_{2}}{t_{1}t_{2}}~\ t_{1}^{\Delta_{1}}t_{2}^{\Delta_{2}}~e^{-\left(
t_{1}+t_{2}\right)  ^{2}+4ut_{1}t_{2}}~. 
\label{a300}
\end{equation}
Using the fact that
\begin{eqnarray*}
&&\int\frac{dt_{1}dt_{2}}{t_{1}t_{2}}\,t_{1}^{\Delta_{1}+n}
t_{2}^{\Delta_{2}+n}e^{-\left(  t_{1}+t_{2}\right)^{2}}=\\
&&=\frac{4^{-n}}{2}
\frac{\Gamma\left(\frac{\Delta}{2}\right)  \Gamma\left(\frac{\Delta+1}{2}\right)}
{\Gamma\left(  \Delta\right)  }\,
\frac{\Gamma(\Delta_{1}+n) \Gamma(\Delta_{2}+n)}{\Gamma\left(\frac{\Delta+1}{2}+n\right)  },
\end{eqnarray*}
we may expand the exponential in (\ref{a300}) in powers of $u$ and resum to
obtain
\[
D_{2}( u)  =\pi^{\frac{d}{2}}
\frac{\Gamma\left( \frac{\Delta}{2}\right)}{\Gamma\left(  \Delta\right)  }\,
F\left(\Delta_{1},\Delta_{2},\frac{\Delta+1}{2},u\right)  ~.
\]
Note that, if we choose
\[
\Delta_{1}=\frac{d}{2}+i\nu~,~\ \ \ \ \ \ \ \ \ \ \ \ \ \ \ \ \Delta_{2}%
=\frac{d}{2}-i\nu~,
\]
the expression for $D_{2}(u)  $ is equal to \cite{Paper4}
\[
\frac{4\pi^{d+1}}{\nu^{2}}\frac{\Gamma\left(  1+i\nu\right)  \Gamma\left(
1-i\nu\right)  }{\Gamma\left(  \frac{d}{2}+i\nu\right)  \Gamma\left(  \frac
{d}{2}-i\nu\right)  }\,\Omega_{i\nu}~,
\]
where $\Omega_{i\nu}$ are the radial Fourier functions in $\mathrm{H}_{d+1} $.
Therefore we have that%
\begin{align*}
\Omega_{i\nu}\left(  \mathbf{w}_{1},\mathbf{w}_{2}\right)   &  =\frac{\nu^{2}%
}{4\pi^{d+1}}\frac{\Gamma\left(  \frac{d}{2}+i\nu\right)  \Gamma\left(
\frac{d}{2}-i\nu\right)  }{\Gamma\left(  1+i\nu\right)  \Gamma\left(
1-i\nu\right)  }\times\\
&  \times\int_{\partial\mathrm{H}_{d+1}}\mathbf{~\widetilde{d\mathbf{z}}%
~\ }~\frac{\left\vert \mathbf{w}_{1}\right\vert ^{\frac{d}{2}+i\nu}\left\vert
\mathbf{w}_{2}\right\vert ^{\frac{d}{2}-i\nu}}{\left(  -2\mathbf{z\cdot w}%
_{1}\right)  ^{\frac{d}{2}+i\nu}\left(  -2\mathbf{z\cdot w}_{2}\right)
^{\frac{d}{2}-i\nu}}~.
\end{align*}

\subsection{Three point function $n=3$, $m=3$}

As is well known, there are no cross--ratios in this case and conformal
invariance determines%
\[
D\left(  \mathbf{w}_{1},\mathbf{w}_{2},\mathbf{w}_{3}\right)  =\frac
{D}{\mathbf{w}_{12}^{\frac{1}{2}\left(  \Delta_{1}+\Delta_{2}-\Delta
_{3}\right)  }\mathbf{w}_{13}^{\frac{1}{2}\left(  \Delta_{1}+\Delta_{3}%
-\Delta_{2}\right)  }\mathbf{w}_{23}^{\frac{1}{2}\left(  \Delta_{2}+\Delta
_{3}-\Delta_{1}\right)  }}
\]
up to an overall constant $D$, determined by the integral%
\[
\frac{2\pi^{\frac{d}{2}}}{\Gamma\left(  \Delta_{1}\right)  \Gamma\left(
\Delta_{2}\right)  \Gamma\left(  \Delta_{3}\right)  }\int dt_{1}dt_{2}%
dt_{3}~t_{1}^{\Delta_{1}-1}t_{2}^{\Delta_{2}-1}t_{3}^{\Delta_{3}-1}%
~e^{-t_{1}t_{2}-t_{1}t_{3}-t_{2}t_{3}}~.
\]
The integral is easily evaluated with the change of variables
\begin{equation}
t_{1}=\sqrt{s_{2}s_{3}/s_{1}}~,\ \ \ \ t_{2}=\sqrt{s_{1}s_{3}/s_{2}%
},~\ \ \ \ \ \ \ \ t_{3}=\sqrt{s_{1}s_{2}/s_{3}}~. \label{a500}%
\end{equation}
The volume form $\prod_{i}dt_{i}/t_{i}$ becomes $\frac{1}{2}\prod_{i}%
ds_{i}/s_{i}$, and the integral evaluates to \cite{AdsCFT}%
\begin{equation}
D=\pi^{\frac{d}{2}}\frac{\Gamma\left(  \frac{\Delta_{1}+\Delta_{2}-\Delta_{3}%
}{2}\right)  \Gamma\left(  \frac{-\Delta_{1}+\Delta_{2}+\Delta_{3}}{2}\right)
\Gamma\left(  \frac{\Delta_{1}-\Delta_{2}+\Delta_{3}}{2}\right)  }%
{\Gamma\left(  \Delta_{1}\right)  \Gamma\left(  \Delta_{2}\right)
\Gamma\left(  \Delta_{3}\right)  }~. \label{a600}%
\end{equation}
Note that the integral determining $D$ converges for $\operatorname{Re}%
(\Delta_{i}+\Delta_{j}-\Delta_{k})>0$, which implies $\operatorname{Re}%
\Delta_{i}>0$.

\subsection{Three point function $n=3$, $m=2$\label{app1_32}}

Let us now assume that $\mathbf{w}_{1}$ is in the bulk of the Milne wedge. We
have a single cross--ratio%
\[
u=\frac{-\mathbf{w}_{1}^{2}\mathbf{w}_{23}}{\mathbf{w}_{12}\mathbf{w}_{13}}%
\]
and the full $D$--function takes the form
\[
D\left(  \mathbf{w}_{1},\mathbf{w}_{2},\mathbf{w}_{3}\right)  =
\frac{D_{3}(u)}{\mathbf{w}_{12}^{\frac{1}{2}
\left(  \Delta_{1}+\Delta_{2}-\Delta_{3}\right)  }\mathbf{w}_{13}^{\frac{1}{2}
\left(\Delta_{1}+\Delta_{3}-\Delta_{2}\right)  }\mathbf{w}_{23}^{\frac{1}{2}\left(
\Delta_{2}+\Delta_{3}-\Delta_{1}\right)  }}~,
\]
with $D_{3}(u)$ determined by the integral representation%
\[
\frac{2\pi^{\frac{d}{2}}}{\Gamma\left(  \Delta_{1}\right)  \Gamma\left(
\Delta_{2}\right)  \Gamma\left(  \Delta_{3}\right)  }\int dt_{1}dt_{2}
dt_{3}~t_{1}^{\Delta_{1}-1}t_{2}^{\Delta_{2}-1}t_{3}^{\Delta_{3}-1}
~e^{-t_{1}t_{2}-t_{1}t_{3}-t_{2}t_{3}-u\,t_{1}^{2}}~.
\]
Applying the change of variables (\ref{a500}) we obtain
\begin{align*}
&  \frac{\pi^{\frac{d}{2}}}{\Gamma\left(  \Delta_{1}\right)  \Gamma\left(
\Delta_{2}\right)  \Gamma\left(  \Delta_{3}\right)  }\times\\
&  \times\int\frac{ds_{1}ds_{2}ds_{3}}{s_{1}s_{2}s_{3}}\,s_{1}^{\frac{\Delta
_{2}+\Delta_{3}-\Delta_{1}}{2}}s_{2}^{\frac{\Delta_{1}+\Delta_{3}-\Delta_{2}%
}{2}}s_{3}^{\frac{\Delta_{1}+\Delta_{2}-\Delta_{3}}{2}}e^{-s_{1}-s_{2}%
-s_{3}-u\,\frac{s_{2}s_{3}}{s_{1}}}~.
\end{align*}
If we expand the exponential in powers of $u$, and formally use the integral
$\int ds~s^{a-1}e^{-s}=\Gamma(a)$, analytically continued to
arbitrary values of $a$, we obtain the formal result%
\begin{equation}
D~F\left(  \frac{\Delta_{1}+\Delta_{2}-\Delta_{3}}{2},\frac{\Delta_{1}%
+\Delta_{3}-\Delta_{2}}{2},1-\frac{\Delta_{2}+\Delta_{3}-\Delta_{1}}%
{2},u\right)  ~,\label{a1000}%
\end{equation}
with the constant $D$ given in (\ref{a600}). The computation is, on the other
hand, only partially correct due to the fact that the integral in $s_{1}$ is
evaluated in the region $\operatorname{Re}a<0$. To deduce the correct answer,
we shall first consider the behavior of the integral $D_{3}\left(  u\right)  $
for $u\rightarrow1$. This is achieved by considering the following
configuration $\mathbf{w}_{1}=\left(  1,1,0\right)  $, $\mathbf{w}_{2}=\left(
1,0,0\right)  $, $\mathbf{w}_{3}=\left(  0,1,0\right)  $ which has $u=1$
exactly$\,$. Choosing the parameterization of $\mathrm{H}_{d+1}$ given by
$\mathbf{y}=\frac{1}{r}\left(  1,r^{2}+y^{2},y\right)  $ with $d\mathbf{y}%
=r^{-1-d}dydr$, the integral (\ref{a200}) is proportional to
\[
\int_{0}^{\infty}\frac{dr}{r}\,r^{\Delta-d}\int_{\mathbb{E}^{d}}\frac
{dy}{\left(  1+r^{2}+y^{2}\right)  ^{\Delta_{1}}\left(  r^{2}+y^{2}\right)
^{\Delta_{2}}}~.
\]
We shall assume, as always, that $\operatorname{Re}\Delta>d$,
$\operatorname{Re}\Delta_{3}<\operatorname{Re}\left(  \Delta_{1}+\Delta
_{2}\right)  $ and $\operatorname{Re}\Delta_{2}<\operatorname{Re}\left(
\Delta_{1}+\Delta_{3}\right)  $, which implies $\operatorname{Re}\left(
\Delta_{1}+\Delta_{2}\right)  >d/2$. The $y$--integral is therefore convergent
and can be explicitly evaluated. The above expression becomes%
\[
\pi^{\frac{d}{2}}\int_{0}^{\infty}\frac{dr}{r}~r^{\Delta_{3}-\Delta_{1}%
-\Delta_{2}}~F\left(  \Delta_{1}+\Delta_{2}-\frac{d}{2},\Delta_{1},\Delta
_{1}+\Delta_{2},-\frac{1}{r^{2}}\right)  ~.
\]
Convergence is now clear. At $r=\infty$ the integrand behaves as
$r^{\Delta_{3}-\Delta_{1}-\Delta_{2}-1}$, whereas close to $r=0$ the two
leading behaviors are given by $r^{\Delta-\frac{d}{2}}$ and $r^{\Delta
_{3}+\Delta_{1}-\Delta_{2}}$. It is then clear that the correct choice
replacing (\ref{a1000}) is given by
\begin{equation}
D_{3}\left(  u\right)  =D^{\prime}~F\left(  \frac{\Delta_{1}+\Delta_{2}%
-\Delta_{3}}{2},\frac{\Delta_{1}+\Delta_{3}-\Delta_{2}}{2},\frac{\Delta}%
{2},1-u\right)  ~,\label{a3000}
\end{equation}
where the normalization
\[
D^{\prime}=\pi^{\frac{d}{2}}\frac{\Gamma\left(  \frac{\Delta_{1}+\Delta
_{2}-\Delta_{3}}{2}\right)  \Gamma\left(  \frac{\Delta_{1}+\Delta_{3}%
-\Delta_{2}}{2}\right)  }{\Gamma\left(  \Delta_{1}\right)  \Gamma\left(
\frac{\Delta}{2}\right)  }~
\]
has been fixed by requiring that $\lim_{u\rightarrow0}D_{3}(u)=D$ whenever the condition 
$\operatorname{Re}\left(\Delta_{2}+\Delta_{3}\right)>\operatorname{Re}\Delta_{1}$ holds. 
In the main text, we are especially interested
in the case $\Delta=d=2$ with $\Delta_{2}=\Delta_{3}=\frac{1-i\mu}{2}$. Then
we have that%
\[
D^{\prime}=\,c(-\mu)\, \frac{64\pi^{4}}{1+\mu^{2}}\ .
\]
Using the properties of the hypergeometric function, it is now trivial to show
that (\ref{a3000}) is given by%
\[
\frac{1}{\mu^{2}c(\mu)}\,u^{-\frac{1+i\mu}{2}}
\big(  \phi_{\mu}(u)  +\phi_{-\mu}(u) \big)  ~,
\]
thus showing (\ref{e5000}).

\section{Polynomial Impact Factors\label{app3}}

Let us consider the integral
\[
\int d\mu~V(\mu)  ~\phi_{\mu}(u)  ~,
\]
with
\[
V(\mu) = \frac{8\pi^{3}}{1+\mu^{2}}\,\frac{\Gamma\left(
\sigma-\frac{1}{2}+\frac{i\mu}{2}\right)  \Gamma\left(  \sigma-\frac{1}%
{2}-\frac{i\mu}{2}\right)  }{\Gamma^{2}\left(  \sigma\right)  }~.
\]
For $u>0$, we may close the contour in the region $\operatorname{Im}\mu<0$. The
contribution to the integral comes from the poles at $i\mu=2\left(
\sigma+n\right)  -1$, with $n$ a non--negative integer, so that we obtain the
following sum of residues
\[
\frac{1}{2}\sum_{n\in\mathbb{N}_{0}}\frac{\left(  -\right) ^{n}}{n!}
\,\frac{\Gamma\left(  2\sigma+n-1\right)  }{\Gamma\left(  2\sigma+2n-1\right)
}\frac{\Gamma^{2}(\sigma+n)  }{\Gamma^{2}(\sigma)
}~u^{\sigma+n}~F\left(  \sigma+n,\sigma+n,2\sigma+2n,u\right)\ .
\]
It can be easily checked that the successive powers $u^{\sigma+n}$ for
$n\geq1$ cancel in the above expression, leaving only the initial $n=0$
contribution $u^{\sigma}/2$. We have then obtained that
\[
\int d\mu~V(\mu)  ~\big[ \phi_{\mu}(u)+\phi_{-\mu}(u) \big]  = u^{\sigma}\ ,
\]
as we needed to show.

\section{$\mathcal{N}=4$ SYM Conventions\label{AppN4}}

In this paper, we use  standard conventions for $\mathcal{N}=4$ SYM.
For the convenience of the reader, we quote the most relevant ones. The
bosonic part of the SYM action is
\[
\frac{1}{g_{\mathrm{YM}}^{2}}\int d^{4}\mathbf{x~}\mathrm{Tr}\left(  -\frac
{1}{2}F_{\mu\nu}F^{\mu\nu}-\nabla_{\mu}\phi_{i}~\nabla^{\mu}\phi_{i}+\frac
{1}{2}\left[  \phi_{i},\phi_{j}\right]  ^{2}\right)  ~,
\]
with $F_{\mu\nu}=\partial_{\mu}A_{\nu}-\partial_{\nu}A_{\mu}-i\left[  A_{\mu
},A_{\nu}\right]  $ and $\nabla_{\mu}\phi_{i}=\partial_{\mu}\phi_{i}-i\left[
A_{\mu},\phi_{i}\right]  $. The six adjoint real scalars $\phi_{i}$ and the
gauge field $A_{\mu}$ are written, in the basis of $N^{2}-1$ generators of
$SU\left(  N\right)  $, as $\phi_{i}=\phi_{i}^{a}T^{a}$ and $A_{\mu}=A_{\mu
}^{a}T^{a}$, where we choose the normalization%
\[
\mathrm{Tr}\left(  T^{a}T^{b}\right)  =\frac{1}{2}\delta^{ab}~.
\]
The structure functions $f^{abc}$ are defined as usual as%
\[
\left[  T^{a},T^{b}\right]  =i~f^{abc}~T^{c}~.
\]
Two useful relations are
\begin{align*}
f^{acd}~f^{bcd}  &  =N~\delta^{ab}~,\\
\left(  T^{a}\right)  _{j}^{i}\left(  T^{a}\right)  _{\ell}^{k}  &  =\frac
{1}{2}\left[  \delta_{j}^{k}\delta_{\ell}^{i}-\frac{1}{N}\,\delta_{j}^{i}%
\delta_{\ell}^{k}\right]  ~,
\end{align*}
which imply, for instance, that%

\begin{equation}
\left[  T^{m_{1}},T^{a}\right]  T^{a}=\frac{N}{2}T^{m_{1}}~. \label{ap1000}%
\end{equation}
We define the complex fields $Z,W$ by%
\[
Z=\frac{1}{\sqrt{2}}\left(  \phi_{1}+i\phi_{2}\right)
~,~\ \ \ \ \ \ \ \ \ \ \ \ \ W=\frac{1}{\sqrt{2}}\left(  \phi_{3}+i\phi
_{4}\right)  ~.
\]
with propagator%
\[ 
\langle Z^{a}\left( \mathbf{x}\right) \bar{Z}^{b}\left( \mathbf{y}\right) \rangle 
=\frac{g_{\mathrm{YM}}^{2}}{4\pi^{2}}\frac{\delta^{ab}}{\left( \mathbf{x}%
-\mathbf{y}\right) ^{2}+i\epsilon}~. 
\] 
The gauge field propagator $A_{\mu}^{a}\left(  \mathbf{x}\right)  A_{\nu}%
^{b}\left(  \mathbf{y}\right)  $ in Feynman gauge is also given by the same
expression, with the addition of the spacetime metric $\eta_{\mu\nu}$.

\section{Impact Factor for $\mathrm{Tr}\left(  Z^{L}\right)  $\label{AppImp}}

\begin{figure}
\begin{center}
\includegraphics[width=12.5cm]{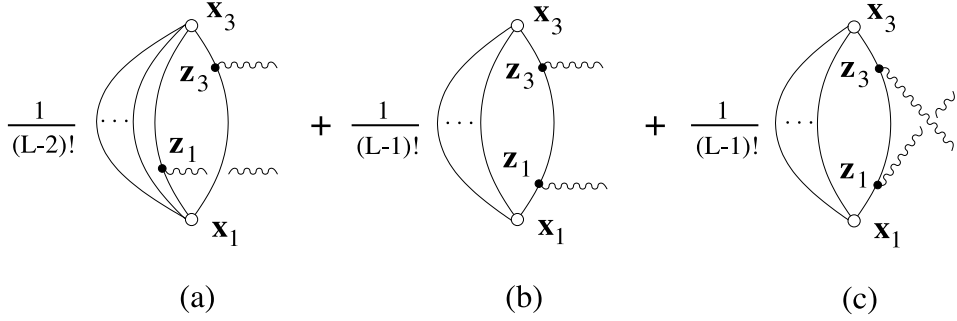}
\end{center}
\caption{Perturbative expansion of the impact factor for the operator
$\mathrm{Tr}\left(  Z^{L}\right)  $. We show explicitly the relevant symmetry
factors associated to the permutations of scalar lines without gluon vertices.}
\label{vertex2}%
\end{figure}

In this appendix, we shall compute the impact factor for the operator%
\[
\mathcal{O}_{1}=c_{L}~\mathrm{Tr}\left(  Z^{L}\right)  ~,~
\]
where the constant $c_{L}$ is fixed by requiring that the $2$--point function
$\left\langle \mathcal{O}_{1}\left(  \mathbf{x}\right)  \mathcal{O}_{1}%
^{\star}\left(  \mathbf{y}\right)  \right\rangle $ be normalized to
$\left\vert \mathbf{x}-\mathbf{y}\right\vert ^{-2L}$. The relevant graphs, to
leading order in the 't Hooft coupling $g^{2}$, are shown in figure
\ref{vertex2}. It is quite clear that the spacetime part of the graphs is
identical to that of graphs in figure \ref{vertex} of section \ref{SecIF} for
the case $L=2$. The only difference comes from the color
structure. To analyze the color factors, we first write the operator
$\mathcal{O}_{1}$ as
\[
\mathcal{O}_{1}=\frac{c_{L}}{L!}\,Z^{a_{1}}\cdots Z^{a_{L}}~\mathrm{T}%
^{a_{1}\cdots a_{L}}~,
\]
where
\[
\mathrm{T}^{a_{1}\cdots a_{L}}~=\sum_{\mathrm{perm}\text{ }\sigma}%
\mathrm{Tr}\left(  T^{a_{\sigma_{1}}}\cdots T^{a_{\sigma_{L}}}\right)  ~.
\]
The coefficient $c_{L}$ is then clearly given by
\[
\frac{c_{L}^{2}}{L!}\,\mathrm{T}^{a_{1}\cdots a_{L}}\mathrm{T}^{a_{1}\cdots
a_{L}}\left(  \frac{g_{\mathrm{YM}}^{2}}{4\pi^{2}}\right)  ^{L}=1~.
\]

Now consider the graphs in figure \ref{vertex2}, starting from the simplest
graphs \ref{vertex2}b,c. In general, the color part is given by
\[
\frac{c_{L}^{2}}{\left(  L-1\right)!}\,f_{m_{1}pa}~f_{pn_{1}b}%
~\mathrm{T}^{m_{1}m_{2}\cdots m_{L}}\mathrm{T}^{n_{1}m_{2}\cdots m_{L}%
}~\left(  \frac{g_{\mathrm{YM}}^{2}}{4\pi^{2}}\right)  ^{L}~.
\]
The above expression is proportional to $\delta_{ab}$ and we may therefore
trace over the indices $a,b$ to obtain the normalization constant%
\[
b_{L}   =-\frac{c_{L}^{2}N}{\left(  L-1\right)  !}\,\mathrm{T}^{m_{1}%
m_{2}\cdots m_{L}}\mathrm{T}^{m_{1}m_{2}\cdots m_{L}}~\left(  \frac
{g_{\mathrm{YM}}^{2}}{4\pi^{2}}\right)  ^{L} =-NL~.
\]
Therefore, the relative contribution of the graphs \ref{vertex2}b,c, compared
to the basic case $L=2$, is given by
\[
\frac{b_{L}}{b_{2}}=\frac{L}{2}~.
\]
Next we analyze the more complex case of graph \ref{vertex2}a. The color part
is given by
\[
\frac{c_{L}^{2}}{\left(  L-2\right)!}\,f_{m_{1}n_{1}a}~f_{m_{2}n_{2}%
b}~\mathrm{T}^{m_{1}m_{2}m_{3}\cdots m_{L}}\mathrm{T}^{n_{1}n_{2}m_{3}\cdots
m_{L}}~\left(  \frac{g_{\mathrm{YM}}^{2}}{4\pi^{2}}\right)  ^{L}~.
\]
Again, we trace over $a,b$ to obtain the normalization constant
\[
a_{L}=\frac{c_{L}^{2}}{\left(  L-2\right)  !}\,f_{m_{1}n_{1}a}f_{m_{2}n_{2}%
a}\mathrm{T}^{m_{1}m_{2}m_{3}\cdots m_{L}}\mathrm{T}^{n_{1}n_{2}m_{3}\cdots
m_{L}}~\left(  \frac{g_{\mathrm{YM}}^{2}}{4\pi^{2}}\right)  ^{L}~.
\]
To compute explicitly the expression above we must compute the expression
$f_{m_{1}n_{1}a}\,f_{m_{2}n_{2}a}\,\mathrm{T}^{m_{1}m_{2}m_{3}\cdots m_{L}}\,
\mathrm{T}^{n_{1}n_{2}m_{3}\cdots m_{L}}$, given by
\begin{align*}
&  L\left(  L-2\right)  !~f_{m_{1}n_{1}a}f_{m_{2}n_{2}a}\sum_{\mathrm{perm}%
\text{ }\sigma}\mathrm{Tr}\left(  T^{m_{\sigma_{1}}}\cdots T^{m_{\sigma_{L}}%
}\right)  \\
&  ~\ \ \ \ \ \ \ \ \ \ \ \ \ \sum_{2\leq j\leq L}\mathrm{Tr}\left(  T^{n_{1}%
}T^{m_{3}}\cdots T^{m_{j}}T^{n_{2}}\cdots T^{m_{L}}\right)\ .
\end{align*}
Substituting $f_{abc}T^{c}\rightarrow-i\left[  T^{a},T^{b}\right]  $ and
performing the sum over $j\,$\ we obtain
\begin{align*}
&  2L\left(  L-2\right)  !~\sum_{\mathrm{perm}\text{ }\sigma}\mathrm{Tr}%
\left(  T^{m_{\sigma_{1}}}\cdots T^{m_{\sigma_{L}}}\right)  \mathrm{Tr}\left(
\left[  T^{m_{1}},T^{a}\right]  T^{a}T^{m_{2}}\cdots T^{m_{L}}\right)  \\
&  =\frac{N}{L-1}\,\mathrm{T}^{m_{1}m_{2}\cdots m_{L}}\mathrm{T}^{m_{1}%
m_{2}\cdots m_{L}}~,
\end{align*}
where we used equation (\ref{ap1000}). We then have that $a_{L}=-~b_{L}$
and that%
\[
\frac{a_{L}}{a_{2}}=\frac{L}{2}~,
\]
thus proving (\ref{imL}).

\end{document}